\documentclass{ws-rv975x65}

\begin{document}

\setcounter{chapter}{0}

\chapter{ANISOTROPIC AND HETEROGENEOUS POLYMERIZED MEMBRANES}

\markboth{L. Radzihovsky}{Anisotropic and Heterogeneous  Polymerized Membranes}

\author{Leo Radzihovsky}

\address{Department of Physics, University of Colorado, Boulder, CO
  80309\\
E-mail: radzihov@colorado.edu}

\begin{abstract}
  In these lectures I describe long scale properties of
  fluctuating polymerized membranes in the presence of 
  network anisotropy and random heterogeneities.
  Amazingly, even infinitesimal amount of these seemingly innocuous
  but physically important ingredients in the membrane's internal
  structure leads to a wealth of striking qualitatively new phenomena.
  Anisotropy leads to a ``tubule'' phase that is intermediate in
  its properties and location on the phase diagram between previously
  discussed ``crumpled'' and ``flat'' phases.  At low temperature, network
  heterogeneity generates conformationally glassy phases, with
  membrane normals exhibiting glass order analogous to spin-glasses.
  The common thread to these distinct membrane phases is
  that they exhibit universal anomalous elasticity, (singularly
  length-scale dependent elastic moduli, universal Poisson ratio,
  etc.), driven by thermal fluctuations and/or disorder and controlled 
  by a nontrivial low-temperature fixed point.
\end{abstract}

\newcommand{\Rv}{{\vec R}}
\newcommand{\rv}{{\mathbf r}}
\newcommand{\ev}{{\hat{e}}}
\newcommand{\hv}{{\vec h}}
\newcommand{\uv}{{\mathbf u}}
\newcommand{\qv}{{\mathbf q}}
\newcommand{\pv}{{\mathbf p}}
\newcommand{\nv}{{\mathbf n}}
\newcommand{\Tr}{{\rm Tr}}
\newcommand{\px}{{\partial_x}}
\newcommand{\py}{{\partial_y}}
\newcommand{\ppi}{{\partial_i}}
\newcommand{\ppj}{{\partial_j}}
\newcommand{\ppa}{{\partial_a}}
\newcommand{\ppb}{{\partial_b}}
\newcommand{\calH}{{\mathcal H}}
\newcommand{\calF}{{\mathcal F}}
\def\mm#1{\underline{\underline{{#1}}}}
\def\um{{\mm{u}}}
\def\Lm{{\mm{\Lambda}}}
\def\Qm{{\mm{Q}}}
\def\Im{{\mm{I}}}
\def\Om{\mm{\mathcal{O}}}
\def\O{{\mathcal{O}}}

\section{Preamble}
\label{intro}

The nature of a membrane's in-plane order, with three well-studied
universality classes, the isotropic liquid\cite{liquid_membrane}, 
hexatic liquid\cite{hexatic_membrane} and solid\cite{NP},
crucially affects its conformational properties.  The most striking
illustration of this (discussed in lectures by Yacov Kantor and by
David Nelson) is the stabilization in polymerized membranes (but not
fluid ones, that are always crumpled\cite{liquid_membrane} beyond a
persistence length\cite{persistent_length}) of a ``flat"
phase\cite{NP}, with long-range orientational order in the local
membrane normals\cite{KKN,PKN}, that is favored at low temperature
over the entropically preferred high-temperature crumpled state.  In a
beautiful ``order from disorder'' phenomenon, a subtle interplay of
wild thermal fluctuations with nonlinear membrane elasticity (made
possible by a finite shear modulus of a solid) infinitely enhances
membrane's bending rigidity, thereby stabilizing the flat phase
against these very fluctuations.\cite{NP}  A universal fluctuation-driven
``anomalous elasticity'' characterizes the resulting flat phase, with
length-scale dependent elastic moduli, non-Hookean stress-strain
relation, and a universal negative Poisson ratio.\cite{AL,DG,LRscsa}

Given such qualitative distinction between liquid and solid membranes,
it is perhaps not too surprising that other in-plane orders can have
important qualitative effects on membrane's long-scale
properties. In these lectures I will discuss 
two such ingredients, namely, membrane network in-plane {\em
  anisotropy} (Section 2) and {\em heterogeneity} (Section 3) in 
fluctuating {\em polymerized} membranes, and will 
show that these seemingly innocuous generalizations, indeed lead to a
wealth of new phenomena that are the subject of these lectures.




\section{Anisotropic Polymerized Membranes}
\label{anisotropy}

\subsection{Motivation and introduction}

\begin{figure}[bth]
  \centering \setlength{\unitlength}{1mm}
\begin{picture}(150,110)(0,0)
\put(0,-40){\begin{picture}(150,0)(0,0)
\includegraphics{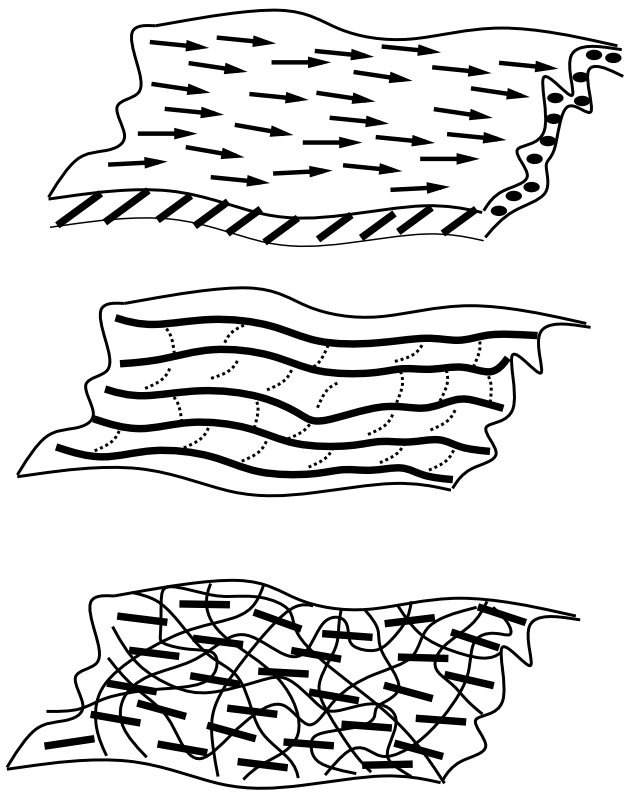}
\end{picture}}
\end{picture}
\caption{Examples of anisotropic polymerized membranes with (a)
  a polymerized-in lipid tilt order in a bilayer membrane, (b) 
  a weakly crosslinked array of linear polymers, and (c) a spontaneous 
  in-plane nematic order in a nematic elastomer membrane.}
\label{anisotropic_membranes}
\end{figure}

In addition to the basic theoretical motivation, the interest in
anisotropic polymerized membranes is naturally driven by a number of
possible experimental realizations, some of which are illustrated in 
Fig.\ref{anisotropic_membranes}.  
One example is a tethered membrane made through photo-polymerization of
a fluid phospholipid membrane exhibiting lipid tilt order.
On average the lipids are tilted
relative to the membrane normal, inducing a vector in-plane order and
an intrinsic elastic anisotropy that can in principle be
aligned with an electric or a magnetic field and polymerized in.
Another possible method\cite{Bensimon} of fabricating polymerized
sheets is by cross-linking a stretched out, aligned array of linear
polymers, that would clearly lead to an intrinsically anisotropic
tethered membrane. One other promising candidate is a two-dimensional
sheet of a nematic elastomer\cite{XingNEmembrane}, a material that
received considerable attention recently because of its novel
elastic and electro-optic properties.\cite{reviewNE}
These cross-linked liquid-crystal polymer gels exhibit all standard
liquid-crystal phases and therefore in their nematic state are highly
anisotropic.

\begin{figure}[bth]
  \centering \setlength{\unitlength}{1mm}
\begin{picture}(150,80)(0,0)
\put(-20,-95){\begin{picture}(150,0)(0,0)
\includegraphics{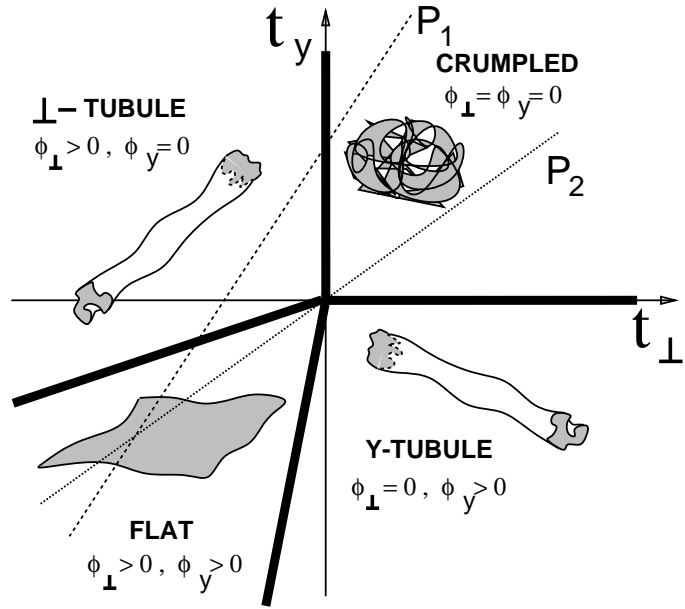}
\end{picture}}
\end{picture}
\caption{Phase diagram for anisotropic tethered
  membranes showing crumpled, tubule and flat
  phases as a function of reduced temperatures $t_\perp\propto
  T-T_c^\perp$, $t_y\propto T-T_c^y$.}
  \label{PhaseDiagram_tubuleF}
\end{figure}

\begin{figure}[bth]
\centering
\setlength{\unitlength}{1mm}
\begin{picture}(150,40)(0,0)
\put(-27,-173){\begin{picture}(150,0)(0,0)
\includegraphics{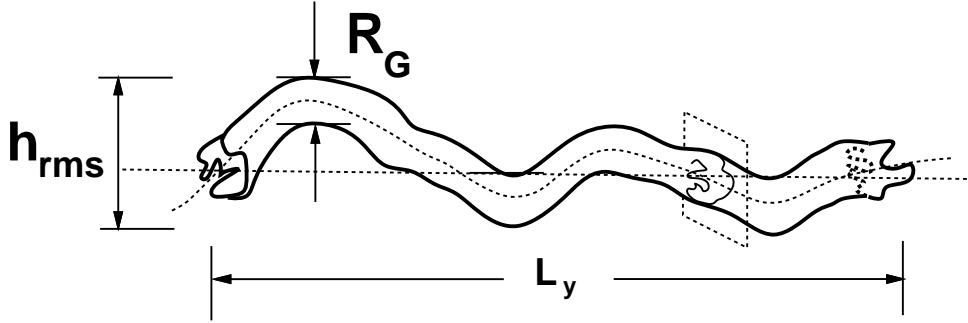}
\end{picture}}
\end{picture}
\caption{Schematic of the y-tubule phase of anisotropic
polymerized membrane, with thickness $R_G$ and
roughness $h_{rms}$.}
\label{tubuleShape}
\end{figure}

Almost 10 years ago, it was discovered\cite{RTtubule} that in-plane
anisotropy has a dramatic qualitative effect on the global phase
diagram of polymerized membranes. As illustrated in
Fig.\ref{PhaseDiagram_tubuleF}, it leads to an entirely new ``tubule'' phase
of a polymerized membrane (see Fig.\ref{tubuleShape}), that is
crumpled along one and extended along the other of the two membrane
axes, with wild undulations about its average cylindrical
geometry.\cite{tubule_tubule}

This is possible because in an anisotropic membrane, where symmetry
between $x_\perp$ and $y$ axes is broken, (e.g., curvature moduli
$\kappa_{x_\perp}\neq\kappa_y$) the $x_\perp$- and $y$-directed tangents
$\vec{\zeta}_\perp=\partial_\perp\vec{r}$ and
$\vec{\zeta}_y=\partial_y\vec{r}$ will generically order at two
distinct temperatures $T_c^\perp\sim\kappa_{x_\perp}$ and $T_c^y\sim\kappa_y$,
respectively, thereby allowing two intermediate tubule states: (1)
$x_\perp$-extended tubule with $\langle\vec{\zeta}_\perp\rangle\neq0$,
$\langle\vec{\zeta}_y\rangle=0$, for $T_c^y < T < T_c^\perp$ and (2)
y-extended tubule with $\langle\vec{\zeta}_\perp\rangle=0$,
$\langle\vec{\zeta}_y\rangle\neq0$, for $T_c^\perp < T < T_c^y$.

The direct crumpling transition occurs in such more
generic anisotropic membrane only for that special set of cuts through
the phase diagram (like P$_2$) that pass through the origin and is in
fact tetra-critical.  Generic paths (like P$_1$) will experience {\it
  two} phase transitions, crumpled-to-tubule, and tubule-to-flat, that
are in distinct universality classes. The tubule phase is therefore not
only generically possible, but actually unavoidable, in membranes with
any type or amount of {\it intrinsic anisotropy}.\cite{notGeneric}


As illustrated in Fig.\ref{tubuleShape} the tubule is characterized by
its thickness $R_G$, (the radius of gyration of its crumpled
boundary), and its undulations $h_{rms}$ transverse to its average
axis of orientation. Quite generally, (for a y-extended tubule of an
$L_\perp\times L_y$ membrane) they obey the scaling laws:
\begin{eqnarray}
R_G(L_\perp,L_y)= L_\perp^{\nu}S_R(L_y/L_\perp^z)\;,\;\;\;\;
h_{rms}(L_\perp,L_y)= L_y^{\zeta}S_h(L_y/L_\perp^z)\;,
\label{hrms}
\end{eqnarray}
where the roughness exponent $\zeta=\nu/z$, and the anisotropy
exponent $z=(1+2\nu)/(3-\eta_\kappa) < 1$ are expressed in terms two
independent universal exponents: the radius of gyration exponent $\nu
<1$ and the anomalous elasticity exponent $\eta_\kappa$ for the tubule
bending rigidity defined by $\kappa\sim L_y^{\eta_\kappa}$.
The {\em universal} scaling functions $S_{R,h}(x)$ have the limiting
forms:
\begin{equation}
S_R(x)\propto
\left\{ \begin{array}{lr}
x^{\zeta-\nu_p/z}, & \mbox{for}\; x\rightarrow 0 \\
\mbox{constant}, & \mbox{for}\; x\rightarrow\infty
\end{array} \right.
\label{fR2d}
\end {equation}
\begin{equation}
\label{fh2d}
S_h(x)\propto
\left\{ \begin{array}{lr}
\mbox{constant}, & \mbox{for}\; x\rightarrow 0 \\
x^{{3\over 2} - \zeta}, & \mbox{for}\; x\rightarrow\infty
\end{array} \right.\;,
\end{equation}
where $\nu_p$ is the 
radius of gyration exponent of a coiled linear
polymer $\approx 3/5$. These asymptotic forms emerge from general
requirements (supported by detailed renormalization group
calculations\cite{RTtubule}) that in the limit of a very thin tubule 
(in the curled up $\perp$ direction) the tubule's height undulations $h_{rms}(L_y)$
reproduce statistics of a linear polymer of length $L_y$ and width
$L_\perp$. And, that in the opposite limit of vanishing $L_y$, scaling
functions must recover the radius of gyration $R_G(L_\perp)\sim
L_\perp^{\nu_p}$ of a linear polymer of length $L_\perp$ and thickness
$L_y$.

The scaling forms, Eq.\ref{fR2d} and \ref{fh2d} imply that for a
"roughly square" membrane -- that is, one with $L_\perp\sim L_y\equiv
L$ -- in the limit $L\rightarrow\infty$
\begin{eqnarray}
R_G(L_\perp\sim L_y\equiv L)\propto L^{\nu}\;,\;\;\;
h_{rms}(L_\perp\sim L_y\equiv L)\propto L^{1-\eta_\kappa z/2}\;.
\label{Rhsq}
\end{eqnarray}
Combining this with detailed 
renormalization group calculations, that find a strictly positive
$\eta_\kappa$,\cite{RTtubule} shows that $h_{rms}<<L$ for a 
large roughly square membrane. Thus, the end-to-end orientational
fluctuations $\theta\sim h_{rms}/L\propto
L^{-\eta_\kappa/2}\rightarrow 0$ as $L\rightarrow\infty$ for such a
roughly square membrane, proving that tubule order (which requires
orientational persistence in the extended direction) {\em is} stable
against undulations of the tubule embedded in $d=3$ dimensions.

The rest of Section \ref{anisotropy} 
is devoted to developing a model of an anisotropic
polymerized membrane and using it to study phase transitions into and
out of the tubule phase and the anomalous elastic properties of a fluctuating
tubule summarized above. 

\subsection{Model}
\label{model}
A model for anisotropic membranes is a generalization of the
isotropic model discussed in David Nelson's lectures.\cite{PKN} As there, we
characterize the configuration of the membrane by the position
${\vec r}({\bf x})$ in the $d$-dimensional embedding space of the
point in the membrane labeled by a $D$-dimensional internal
co-ordinate ${\bf x}$, with $d=3$ and $D=2$ corresponding to the
physical case of interest.

Rotational and translational symmetries require that the
Landau-Ginzburg free energy $F$ is 
an expansion in powers of tangent vectors 
$\vec{\xi}_{\perp,y}=\partial_{\perp,y}{\vec r}$ and their
gradients  
\begin{eqnarray}
&&F[{\vec r}({\bf x})] = {1\over2}\int d^{D-1}x_\perp dy
\bigg[\kappa_\perp\left(\partial_\perp^2\vec{r}\right)^2 +
\kappa_y\left(\partial_y^2\vec{r}\right)^2+
\kappa_{\perp y}\partial_y^2\vec{r}\cdot\partial_\perp^2\vec{r}\nonumber\\ 
&+&
t_\perp\left(\partial^\perp_\alpha\vec{r}\right)^2
+t_y\left(\partial_y\vec{r}\right)^2 +
{u_{\perp\perp}\over2}\left(\partial^\perp_\alpha\vec{r}
\cdot\partial^\perp_{\beta}\vec{r}\right)^2 +
{u_{y y}\over2}\left(\partial_y\vec{r}\cdot\partial_y\vec{r}\right)^2 +
u_{\perp y}\left(\partial^\perp_\alpha\vec{r}
\cdot\partial_y\vec{r}\right)^2 \nonumber\\
&+&
{v_{\perp\perp}\over2}\left(\partial^\perp_\alpha\vec{r}
\cdot\partial^\perp_\alpha\vec{r}\right)^2 +
v_{\perp y}\left(\partial^\perp_\alpha\vec{r}\right)^2
\left(\partial_y\vec{r}\right)^2\bigg]+
{b\over2}\int d^{D}x \int d^{D}x'
\hspace{0.2cm}\delta^{(d)}\big({\vec r}({\bf x})-{\vec r}({\bf x'})\big)\;, 
\nonumber\\
\label{Fc}
\end{eqnarray}
where we have taken the membrane to be isotropic in the $D-1$
membrane directions $\bf{x}_\perp$, orthogonal to one special direction, $y$. 
While at first glance $F$ might appear quite formidable, in fact (aside from
the last term), in terms of the tangent order parameter
$\vec{\xi}_{\perp,y}$ it has a standard form of the Landau's $\phi^4$
theory.  The first three terms in $F$ (the $\kappa$ terms) represent the
anisotropic bending energy of the membrane.  The elastic constants
$t_\perp$ and $t_y$ are the most strongly temperature dependent
parameters in the model, changing sign from large, positive values at
high temperatures to negative values at low temperatures.  
The $u$ and $v$ quartic elastic terms are needed to stabilize the membrane 
when one or both of the elastic constants $t_\perp$, $t_y$ become negative.
The final, $b$ term in Eq.\ref{Fc} represents the self-avoidance of the
membranes, i.e., its steric or excluded volume interaction.

\subsection{Mean-field theory}
\label{mft}

In mean-field theory, we seek a configuration ${\vec r}({\bf x})$ that
minimizes the free energy, Eq.\ref{Fc}.  The curvature energies
$\kappa_\perp\left(\partial_\perp^2\vec{r}\right)^2$ and
$\kappa_y\left(\partial_y^2\vec{r}\right)^2$ are clearly minimized
when ${\vec r}({\bf x})$ is linear in ${\bf x}$
\begin{equation}
{\vec r}({\bf x})=\left(\zeta_\perp {{\bf x}_\perp}, \zeta_y y, 0, 0,
\ldots , 0\right)\;.
\label{ansatz}
\end{equation}
Inserting Eq.\ref{ansatz} into Eq.\ref{Fc}, and for now neglecting the
self-avoidance, we obtain the mean-field free energy density for anisotropic
membranes
\begin{eqnarray}
f_{\rm mft}={1\over2}\big[t_y\zeta_y^2+
t_\perp(D-1) \zeta_\perp^2 +
{1\over2}u'_{\perp\perp}(D-1)^2\zeta_\perp^4
+ {1\over2}u_{y y}\zeta_y^4 + v_{\perp y}(D-1)\zeta_\perp^2
\zeta_y^2\big]\;,\nonumber\\
\label{Fzeta}
\end{eqnarray}
Minimizing the free energy over order parameters $\zeta_\perp$ 
and $\zeta_y$ yields two possible phase diagram topologies,
depending on whether $u_{\perp\perp}' u_{y y} > v_{\perp y}^2$ 
or $u_{\perp\perp}' u_{y y} < v_{\perp y}^2$.\cite{NFAM}

For $u_{\perp\perp}' u_{y y} > v_{\perp y}^2$, the phase
diagram is given in Fig.\ref{PhaseDiagram_tubuleF}.  For $t_\perp>0$
and $t_y>0$, average tangent vectors $\zeta_\perp$ and $\zeta_y$ both
vanish, describing a crumpled (collapsed to the origin) membrane.

In the regime between the positive $t_\perp$-axis 
and the $t_y < 0$ part of the $t_y=(u_{y y}/v_{\perp y})t_\perp$ line, 
lies the y-tubule phase, characterized by $\zeta_\perp=0$ 
and $\zeta_y=\sqrt{|t_y|/u_{y y}}>0$:
the membrane is extended in the y-direction but crumpled in all $D-1$
$\perp$-directions.

The $\perp$-tubule phase is the analogous phase with the $y$ and
$\perp$ directions reversed, $\zeta_y=0$ and
$\zeta_\perp=\sqrt{|t_\perp|/u_{\perp\perp}}>0$, and lies between
the $t_\perp < 0$ segment of the line $t_y=(v_{\perp
y}/u'_{\perp\perp}) t_\perp$ and the positive $t_y$ axis. 

Finally, the flat phase, characterized by both 
\begin{eqnarray}
\zeta_\perp&=&\left[(|t_\perp| u_{y y}-|t_y| v_{\perp y})
/(u_{\perp\perp}' u_{y y} - v_{\perp y}^2)\right]^{1/2} > 0\;,
\label{zeta_perp}\\
\zeta_y&=&\left[(|t_y| u_{\perp\perp}-|t_\perp| v_{\perp y})
/(u_{\perp\perp}' u_{y y}-v_{\perp y}^2)\right]^{1/2} > 0\;,
\label{zeta_y}
\end{eqnarray}
lies between the $t_\perp < 0$ segment of the line $t_y=(u_{y
y}/v_{\perp y})t_\perp$ and the $t_y < 0$ segment of the line
$t_y=(v_{\perp y}/u'_{\perp\perp}) t_\perp$.

\begin{figure}[bth]
\centering
\setlength{\unitlength}{1mm}
\begin{picture}(150,80)(0,0)
\put(-20,-100){\begin{picture}(150,0)(0,0)
\includegraphics{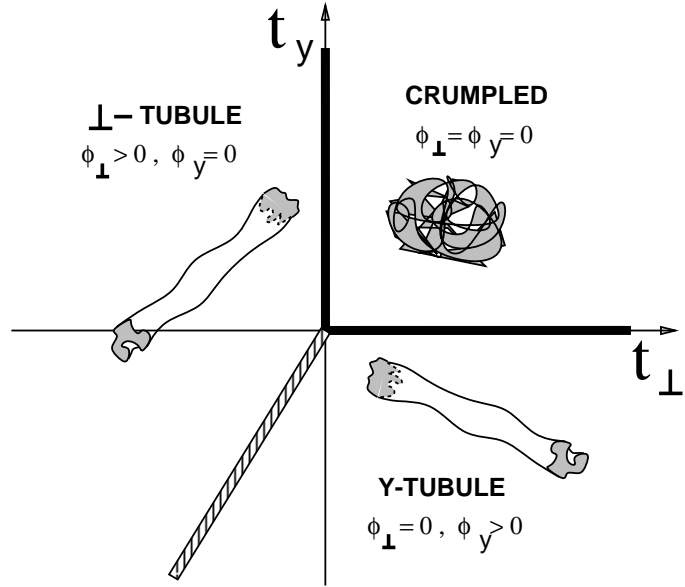}
\end{picture}}
\end{picture}
\caption{Phase diagram for anisotropic tethered membranes showing a 
new tubule phase, for the range of elastic parameters when the 
intermediate flat phase disappears. A first-order phase transition 
separates $y$- and $\perp$-tubule phases.}
\label{PhaseDiagram_tubuleNF}
\end{figure}

For $u_{\perp\perp}' u_{y y} < v_{\perp y}^2$, the flat phase
disappears, and is replaced by a direct first-order transition from
$\perp$-tubule to $y$-tubule along the locus $t_y=(v_{\perp
y}/u'_{\perp\perp}) t_\perp$ (see Fig.\ref{PhaseDiagram_tubuleNF}) .
The boundaries between the tubule and the crumpled phases remain the
positive $t_y$ and $t_\perp$ axes, as for $u_{\perp\perp}' u_{y y} >
v_{\perp y}^2$ case.

\subsection{Fluctuations and self-avoidance in the crumpled and flat
phases}
\label{flat_and_crumpled}
Although anisotropy has dramatic effects on the phase diagram of
polymerized membranes, it does not modify the nature of the crumpled
and flat phases.\cite{EfieldComment}
To see this for the crumpled phase note that when 
both $t_\perp$ and $t_y$ are positive, all of the other local terms 
in Eq.\ref{Fc}, i.e., the $\kappa$, $u$, and $v$-terms, are
subdominant at long wavelengths. When these irrelevant terms are
neglected, a simple change of variables
${\bf x}_\perp={\bf x}'\sqrt{t_\perp/t_y}$ makes the remaining energy
isotropic.  Thus, the entire crumpled phase is identical in its
scaling properties to that of isotropic membranes.
In particular, the membrane in this phase has a radius of gyration
$R_G(L)$ which scales with membrane linear dimension  $L$  like
$L^\nu$, with $\nu=(D+2)/(d+2)$ in Flory theory, and very similar
values predicted by $\epsilon$-expansion
techniques.\cite{KNsa,KKNsa,ALsa}

Fluctuations in the flat phase can be incorporated by
considering small deviations from the mean-field conformation
in Eq.\ref{ansatz}
\begin{equation}
{\vec r}({\bf x})=\left(\zeta_\perp{\bf x_\perp}+{u_\perp}({\bf x}),
\zeta_
y y +u_y({\bf x}), {\vec h}({\bf x})\right)\;,
\label{fluct}
\end{equation}
where $u_\alpha({\bf x})$ are $D$ in-plane phonon fields and 
$h_i({\bf x})$ are $d_c=d-D$ out-of-plane height undulation fields.
Inserting this into free energy, Eq.\ref{Fc}, with
$t_\perp$ and $t_y$ both in the range in which the flat phase is
stable, we obtain the uniaxial elastic energy studied by 
Toner.\cite{TonerAnisotropy} As he showed, amazingly, fluctuation
renormalize this anisotropic elastic energy into the 
{\it isotropic} membrane elastic energy studied previously.\cite{NP,AL,DG,LRscsa} 
Therefore, in the flat phase, and at sufficiently long
scales, the anisotropic membranes behave exactly like isotropic ones.

\subsubsection{Anomalous elasticity of the flat phase}
\label{flatAnomElast}

This in particular implies that the flat phase of
anisotropic membranes is stable against thermal fluctuations
even though it breaks continuous (rotational) symmetry and 
is two-dimensional.\cite{MerminWagner}
As in isotropic membranes, this is due to the fact that at long
wavelengths these very thermal fluctuations drive the effective
(renormalized) bend modulus 
$\kappa$ to infinity\cite{NP,AL,DG,LRscsa}, thereby suppressing effects of
these same fluctuations that seek to destabilize the flat phase,
resulting in an Esher-like ``order-from-disorder'' phenomenon.

Specifically, $\kappa(q)$ becomes wavevector dependent, and 
diverges like $q^{-\eta_\kappa}$ as $q\rightarrow 0$.  In the flat
phase the standard Lam\'e coefficients $\mu$ and $\lambda$\ \
\cite{LandauLifshitz} are also infinitely renormalized and become
wavevector dependent, vanishing in the $q\rightarrow0$ limit as
$\mu(q)\sim\lambda(q)\sim q^{\eta_u}$. The flat phase is furthermore
novel in that it is characterized by a universal {\em negative}
Poisson ratio\cite{AL,LRscsa} which for $D=2$ is defined as the long
wavelength limit $q\rightarrow0$ of
$\sigma=\lambda(q)/(2\mu(q)+\lambda(q))$. The transverse undulations
in the flat phase, i.e. the membrane roughness $h_{rms}$ scales with
the internal size of the membrane as $h_{rms}\sim L^\zeta$, with
$\zeta=(4-D-\eta_\kappa)/2$, exactly. Furthermore, an underlying
rotational invariance imposes an exact Ward identity between
$\eta_\kappa$ and $\eta_u$, $\eta_u+2\eta_\kappa=4-D$. This leaves only a
single independent exponent, characterizing the properties
of the flat phase of even anisotropic membranes.

\begin{figure}[!htbp]
\centering
\setlength{\unitlength}{1mm}
\begin{picture}(150,25)(0,0)
\put(-65,-125){\begin{picture}(150,0)(0,0)
\includegraphics{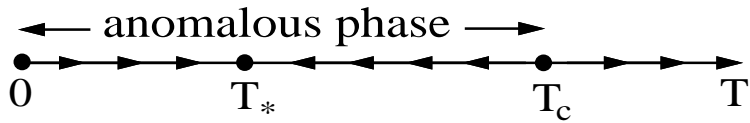}
\end{picture}}
\end{picture}
\caption{Renormalization group flow for anomalously elastic
  solids, with $T_c$ the transition temperature to the ordered state
  and $T_*$ a nontrivial infrared stable fixed point controlling
  properties of the strongly interacting ordered, critical-like state.}
\label{softsolidRGflow}
\end{figure}

To appreciate how exotic and unusual this anomalous elasticity really
is one only needs to observe that most ordered and disordered states
of matter (systems with quenched disorder being prominent exceptions
\cite{randomFP}), 
are in a sense trivial, with fluctuations about them described by a 
harmonic theory controlled by a Gaussian fixed point.\cite{commentGaussianFP}  
That is, generically, qualitatively important effects of fluctuations
are confined to the vicinity of isolated critical points, where a system
is tuned to be ``soft'', and characterized by low energy excitations. 
However, there exists a novel class of systems, with flat phase of 
polymerized membranes as a prominent member (that also includes
smectic\cite{GP,RTsmectic} and columnar liquid
crystals\cite{RTcolumnar}, vortex lattices in putative magnetic
superconductors\cite{RTmsc}, and nematic 
elastomers\cite{reviewNE,GL,LMRX,NEelXR,NEelSL,NEprlXR})
whose ordered states are a striking exception to this rule. A unifying
feature of these phases is their underlying, spontaneously broken
rotational invariance, that strictly enforces a particular
``softness'' of the corresponding Goldstone mode Hamiltonian. As a
consequence, the usually small nonlinear elastic terms are in fact
comparable to harmonic ones, and therefore must be taken into account.
Similarly to their effects near continuous phase transitions (where
they induce universal power-law anomalies), but
extending throughout an ordered phase, fluctuations drive
nonlinearities to qualitatively modify such soft states.  The
resulting strongly interacting ordered states at long scales exhibit
rich phenomenology such as a universal nonlocal elasticity, a strictly
nonlinear response to an arbitrarily weak perturbation and a universal
ratio of wavevector-dependent singular elastic moduli, all controlled
by a nontrivial infrared stable fixed point illustrated in
Fig.\ref{softsolidRGflow}.

\subsubsection{SCSA of the flat phase}
\label{SCSAflat}

The study of such anomalous elasticity in polymerized membranes was
initiated by Nelson and Peliti using a simple one-loop self-consistent
theory, that assumed a non-renormalization of the in-plane elastic Lame'
constants.\cite{NP} They found that phonon-mediated 
interaction between capillary waves leads to a divergent bending
rigidity with $\eta_\kappa=1$ and membrane roughness exponent
$\zeta=1/2$. Subsequent controlled  $\epsilon=4-D$ and
$1/d_c$ expansions\cite{AL,DG} confirmed existence of anomalous elasticity, and 
discovered an additional effect of softening (screening) of in-plane
elasticity by out-of-plane undulations, that lead to a vanishing of
long-scale Lame' coefficients $\lambda(q) \sim \mu(q) \sim
q^{\eta_u}$. 


Here we treat anomalous elasticity within the so-called
self-consistent screening approximation (SCSA), first applied to the
study of the flat phase of polymerized membranes by Le Doussal and 
Radzihovsky\cite{LRscsa}. The
attractive feature of SCSA is that it is becomes exact in three
complementary limits. By construction, it is exact in the large
embedding dimension $d_c\rightarrow\infty$ limit and agrees with the
systematic $1/d$ analysis of Guitter, et al.\cite{DG}
Because of Ward identities
associated with rotational invariance it is exact (at any $d$) to lowest
order $\epsilon=4-D$, i.e., agrees with one-loop results of Aronovitz and 
Lubensky.\cite{AL} (It would be very interesting to check predictions
of SCSA to order $\epsilon^2$, however technical difficulties with
two-loop calculations have so far precluded
this\cite{RJ2loop}). And finally it gives exact value of 
$\eta_\kappa(D)$ for $d_c=0$. Given these exact constraints, it is not 
surprising that for physical dimensions ($D=2$, $d=3$) SCSA exponents
and the universal negative Poisson ratio\cite{LRscsa} 
compare so well with latest, largest scale numerical simulations, discussed
in lectures by Mark Bowick.\cite{zetaBowick}

As discussed above, at sufficiently long scales, flat phase of
an anisotropic polymerized membrane is identical to that of an
isotropic one\cite{TonerAnisotropy}, and is described by an isotropic
effective free energy that is a sum of a bending and an in-plane
elastic energies:
\begin{equation}
F[\vec{h},{\bf u}]= \int d^Dx  [ {\kappa \over 2} (\nabla^2\vec{h})^2
    + \mu u_{\alpha \beta}^2 +{\lambda \over 2} u_{\alpha \alpha}^2 ]\;,
\label{Fflat}
\end{equation}
where the strain tensor is $ u_{\alpha \beta}=
 {1\over 2} ( {\partial _{\alpha} u_{\beta}} +
{\partial _{\beta} u_{\alpha} } +
{\partial _{\alpha} {\vec h}}\cdot{\partial_{\beta} {\vec h}})$.
To implement the SCSA in the flat phase it is convenient to first 
integrate out the noncritical phonons fields $u_\alpha$
obtaining a quartic theory for interacting Goldstone tangent vector 
modes ${\partial_\alpha\vec h}$, described by free energy
\begin{equation}
F[\vec{h}]= \int d^Dx  [{\kappa \over 2} (\nabla^2\vec{h})^2
+{1 \over {4 d_c} }(\partial_\alpha\vec{h}\cdot\partial_\beta\vec{h}) 
R_{\alpha \beta, \gamma \delta}
(\partial_\gamma\vec{h}\cdot\partial_\delta\vec{h})]
\end{equation}
where for convenience, we rescaled Lam\'e coefficients
so that the quartic coupling is of order $1/d_c$.
The four-point coupling fourth-rank tensor is given by
\begin{equation} 
R_{\alpha \beta, \gamma \delta}={K-2\mu\over2(D-1)} P^T_{\alpha \beta} 
P^T_{\gamma \delta}+{\mu\over2}\left(P^T_{\alpha \gamma} P^T_{\beta \delta } 
+ P^T_{\alpha \delta} P^T_{\beta \gamma}\right)\;,
\label{Rdefine}
\end{equation}
where $P^T_{\alpha \beta}=\delta_{\alpha \beta} -
q_{\alpha}q_{\beta}/q^2$ is a transverse (to ${\bf q}$) 
projection operator. The convenience of this decomposition
is that $K= 2\mu ( 2 \mu + D \lambda)/(2 \mu + \lambda)$ and 
$\mu$ moduli renormalize independently and multiplicatively.

To determine the renormalized elasticity, we set up a 
$1/d_c$-expansion\cite{largeN,largeNcomment,ZinnJustin} for 2-point and 4-point
correlation functions of $\vec h$, and turn them into a closed
self-consistent set of 
two coupled integral equations for self-energy $\Sigma({\bf k})$ that
define SCSA
\begin{eqnarray}
\Sigma({\bf k}) &=& {2 \over d_c} k_{\alpha}k_{\beta}k_{\gamma}k_{\delta}
\int_q \tilde{R}_{\alpha \beta, \gamma \delta}({\bf q}) G({\bf k}-{\bf
  q})\;,\label{SigmaEqn}\\
\tilde{R}_{\alpha \beta, \gamma \delta}({\bf q})&= &
R_{\alpha \beta, \gamma \delta}({\bf q}) - 
R_{\alpha \beta, \mu \nu}({\bf q})
\Pi_{\mu\nu,\mu'\nu'}({\bf q})\tilde{R}_{\mu'\nu',\gamma \delta}({\bf q})\;,
\label{SCSAeqn1}
\end{eqnarray}
where $G({\bf k})\equiv 1/(\kappa k^4 +
\Sigma({\bf k}))\equiv1/(\kappa({\bf k})k^4)$ is the renormalized
propagator, $R_{\alpha \beta, \gamma \delta}({\bf q})$ the bare
quartic interaction vertex and 
$\tilde{R}_{\alpha \beta, \gamma \delta}({\bf q})$ 
the "screened" interaction vertex dressed by the 
vacuum polarization bubbles $\Pi_{\alpha \beta, \gamma \delta}({\bf q})
=\int_p p_{\alpha} p_{\beta} p_{\gamma} p_{\delta} G({\bf p}) G({\bf
  q}-{\bf p})$, and tensor multiplication is defined above. 
%

In the long wavelength limit these integral equations are solved
exactly by a renormalized propagator $G({\bf k})\approx 1/\Sigma({\bf k}) 
\approx Z/ k^{4-\eta_\kappa}$, with $Z$ a non-universal amplitude.
Substituting this ansatz into Eqs.\ref{SigmaEqn},\ref{SCSAeqn1} and solving for the
renormalized elastic Lam\'e moduli, we find that indeed they must
vanish as a universal power, with $\mu({\bf q})\propto\lambda({\bf
  q})\sim q^{\eta_u}$, and the phonon anomalous exponent 
$\eta_u=4-D-2\eta_\kappa$ related to $\eta_\kappa$ by
dimensional analysis (power-counting on $q$) first obtained by Nelson 
and Peliti.\cite{NP} This recovers the celebrated exponent relation 
enforced by the underlying rotational invariance of the membrane 
in the embedding space. The information about remaining independent 
exponent $\eta_\kappa$ resides in the 
$\eta_\kappa$-dependent {\em amplitudes} of above equations. Cancelling out
the nonuniversal scale $Z$, we obtain
\begin{equation}
d_c={2 \over \eta_\kappa} 
D(D-1) { { \Gamma[1+{1\over 2}\eta_\kappa] \Gamma[2-\eta_\kappa] 
\Gamma[\eta_\kappa+D] \Gamma[2-{1\over 2}\eta_\kappa] }
\over {\Gamma[{1\over 2}D + {1\over 2}\eta_\kappa] 
\Gamma[2-\eta_\kappa-{1\over 2}D] 
\Gamma[\eta_\kappa+{1\over 2}D] \Gamma[{1\over 2}D+2-{1\over 2}\eta_\kappa]}},
\label{etaDd}
\end{equation}
that determines $\eta_\kappa(D,d)$ within SCSA.  
For $D=2$ this equation can be simplified, and one finds
\begin{equation}
\eta_\kappa(D=2,d_c)={ 4 \over {d_c + \sqrt{16 - 2 d_c + d_c^2}}}\;.
\label{eta2d}
\end{equation}

Thus for physical membranes ($d_c=1$) 
we obtain: $\eta_\kappa=0.821$, $\eta_u=0.358$ and:
\begin{equation}
\zeta=1-{\eta_\kappa \over 2}={ \sqrt{15} - 1 \over{ \sqrt{15} + 1}} = 0.590\;.
\label{zeta}
\end{equation}
At long length scales SCSA also gives a universal ratio between 
renormalized in-plane elastic moduli determined by $\tilde{R}_{\alpha
  \beta, \gamma \delta}({\bf q})$, Eq.\ref{SCSAeqn1}, and therefore 
predicts a universal and {\em negative} Poisson ratio 
\begin{equation}
\mathop{{\rm Lim}}\limits_{q\rightarrow 0}\;\;
\sigma\equiv{\lambda(q)\over
2\mu(q)+(D-1)\lambda(q)} =-{1\over 3}\;.
\label{PoissonRatio}
\end{equation}
that compares extremely well with the most recent and largest 
simulations.\cite{zetaBowick}

Expanding the result Eq.\ref{etaDd} in $1/d_c$ we obtain:
\begin{eqnarray}
\eta_\kappa&=&{8 \over d_c} {D-1 \over D+2} {\Gamma[D] 
\over {\Gamma[{D \over 2}]^3 \Gamma[2-{D \over 2}]}}+O({1 \over
  {d_c^2}})\;,\\
&=& {2 \over d_c}+O({1 \over {d_c^2}})\ ,\ \ \  {\mbox{for}}\  D=2\;,
\label{etaLarge_d}
\end{eqnarray}
which coincides with the exact result\cite{DG,AL}, as expected by 
construction of the SCSA.
Similarly, expanding Eq.\ref{etaDd} 
to first order in $\epsilon=4-D$ we find: 
\begin{equation}
\eta_\kappa={\epsilon \over {2 + {d_c/12}}}\;,
\label{etaD}
\end{equation}
also in agreement with the exact result.\cite{AL} 
This is not, however, 
a general property of SCSA, but is special to membranes, and can be 
traced to the convergence of the vertex and box diagrams.

Because in the flat phase, widely intrinsically separated parts of
the membranes (i.e., points ${\bf x}$ and ${\bf x'}$, with $|{\bf
x}-{\bf x'}|$ large) do not bump into each other (i.e., never have
${\vec r}({\bf x})={\vec r}({\bf x'})$), the self-avoidance
interaction in Eq.\ref{Fc} is irrelevant in the flat phase. Hence we
expect above predictions to accurately describe conformational
properties of physical isotropic and anisotropic polymerized membranes.

\subsection{Fluctuations in ``phantom'' tubules}
\label{tubuleF}

Thermal fluctuations of the (y-) tubule about the mean-field state
${\vec r}_o({\bf x})=\zeta_y\big(y, \vec 0\big)$ are described by conformation,
\begin{equation}
{\vec r}({\bf x})=
\big(\zeta_y y + u({\bf x}),{\vec h}({\bf x})\big)\;,
\label{tubulecoordinate}
\end{equation}
where ${\vec h}({\bf x})$ is a $d-1$-component vector field orthogonal to
the tubule and $u({\bf x})$ is a scalar phonon field along
the tubule (taken along $y$-axis). The order parameter $\zeta_y$ is a
tubule extension scale  that is slightly modified by thermal
fluctuations from the mean-field value given in Sec.\ref{mft} and
is determined by the condition that all linear terms in ${\vec h}({\bf
x})$ and $u({\bf x})$ in the renormalized elastic free energy exactly
vanish.  This criterion guarantees that
${\vec h}({\bf x})$ and $u({\bf x})$ represent fluctuations around the
true tubule ground state.

Inserting the decomposition Eq.\ref{tubulecoordinate} into the free
energy Eq.\ref{Fc}, neglecting irrelevant terms (e.g., phonon
nonlinearities), and, for the moment
ignoring the self-avoidance interaction, gives
$F=F_{mft}+F_{el}$, where $F_{\rm mft}$ is given by Eq.\ref{Fzeta} (with
$\zeta_\perp = 0$ and 
$F_{el}[u({\bf x}),{\vec h}({\bf x})]$ is the fluctuating elastic
free energy part
\begin{eqnarray}
F_{el}={1\over2}\int d^{D-1}x_\perp dy
\bigg[\kappa(\partial^2_y{\vec h})^2 + t(\partial^\perp_\alpha \vec h)^2
+ g_\perp(\partial^\perp_\alpha u)^2+
g_y\big(\partial_y u+{1\over2}(\partial_y{\vec h})^2\big)^2\bigg],\;\;\;\;\;\;
\label{elFtubule}
\end{eqnarray}
where $\kappa\equiv\kappa_y$, $t\equiv t_\perp +v_{\perp y}\zeta_y^2$,
$g_y\equiv u_{yy}\zeta_y^2/2$, $g_\perp\equiv t + u_{\perp
y}\zeta_y^2$, and $\gamma=t_y + u_{y y}\zeta_y^2$ are elastic constants.

The underlying rotational invariance of the tubule in the embedding
space is responsible for two essential features of $F_{el}$. Firstly,
it enforces a strict vanishing of the $(\partial_y{\vec h})^2$ 
tension-like term, with curvature $(\partial_y^2 {\vec h})$
as the lowest order harmonic term in the Goldstone tangent mode 
$\partial_y\vec h$. The result is highly anisotropic bulk elastic
energy.  Rotational invariance also ensures that nonlinear
elasticity can only come in through powers of nonlinear strain tensor 
$E(u,{\vec h})\equiv \partial_y u+{1\over2}(\partial_y{\vec h})^2$, and 
that this property must be preserved upon renormalization.

\subsubsection{Anomalous elasticity of the tubule phase}
\label{tubuleAnomalous}

Within harmonic approximation tubule, {\em bulk} rms transverse
height undulations are given by
\begin{eqnarray}
\langle|\vec{h}({\bf x})|^2\rangle
&\approx&\int_{q_\perp>L_\perp^{-1}}{d^{D-1} q_\perp d q_y\over(2\pi)^D}
{1\over t q_\perp^2 + \kappa q_y^4}\propto L_\perp^{5/2-D}\;.\nonumber\\
\label{hfluct}
\end{eqnarray}
This suggests that for ``phantom'' tubules, the upper critical dimension 
$D_{uc}=5/2$, contrasting with $D_{uc} = 4$ for the flat phase.\cite{AL,DG}
This also implies that rms fluctuations of the tubule normals are given by
\begin{equation}
\langle|\delta n_y({\bf x})|^2\rangle\propto L_\perp^{3/2-D}\;.
\label{Dlc}
\end{equation}
Since this diverges in the infra-red $L_\perp\rightarrow\infty$
for $D\leq D_{lc}=3/2$, this harmonic bulk mode  analysis (ignoring
anomalous elasticity and zero modes) suggests
that the lower critical dimension $D_{lc}$ below which the tubule is
necessarily crumpled is given by $D_{lc}=3/2$.

As for the flat phase, one needs to assess the role of elastic
nonlinearities that appear in free energy $F_{el}$, Eq.\ref{elFtubule}. 
One way to do this is to integrate out the phonon $u$ (which, at long
scales can be done exactly). This leads to the only remaining
nonlinearity in $\vec h$
\begin{equation}
F_{\rm anh}[\vec{h}]= {1\over4}\int d^{D-1}x d y 
(\partial_y\vec{h}\cdot\partial_y\vec{h}) 
V_h (\partial_y\vec{h}\cdot\partial_y\vec{h})],
\label{Fanh}
\end{equation}
with the Fourier transform of the kernel given by
\begin{equation}
V_h({\bf q})={g_y g_\perp q_\perp^2\over g_y q_y^2+g_\perp q_\perp^2}\;.
\label{vertexV_h}
\end{equation}
Because of the $|{\bf q}|_\perp\approx q_y^2$ ($z=1/2$) 
anisotropy of the bulk $\vec h$ modes, $V_h({\bf q})$ scales as
$g_\perp q_\perp$ at long scales, and therefore is strongly irrelevant
near the Gaussian fixed point as long as $g_\perp$ is not renormalized
(but see below). It is straightforward to verify to {\it all} orders 
in perturbation theory, that in a phantom tubule, 
there is no such renormalization of $g_\perp$.\cite{RTtubule} 

However, as asserted earlier, the {\it full} elasticity Eq.\ref{elFtubule},
{\it before} $u$ is integrated out, {\it is} anomalous, because 
$g_y({\bf q})$ is driven to zero as $q_y\rightarrow 0$, according to 
\begin{equation}
g_y({\bf q})=q_y^{\eta_u} S_g(q_y/q_\perp^{z})\;,\label{gyAnsatz}
\end{equation}
with $S_g(x)$ universal scaling function:
\begin{eqnarray}
S_g(x)&\propto&\left\{\begin{array}{lr}
\mbox{constant}, & x\rightarrow\infty\\
x^{-\eta_u}, & x\rightarrow 0\;,
\label{fg_limits}
\end{array} \right.
\end{eqnarray}
and exact exponents:
\begin{eqnarray}
z={1\over2}\;,\;\;\eta_u= 5-2D\;.
\label{z_eta}
\end{eqnarray}
One simple way to see this is to note that rotational invariance
enforces graphical corrections to preserve the form of the nonlinear 
strain tensor 
$E(u,{\vec h})\equiv \partial_y u+{1\over2}(\partial_y{\vec h})^2$. This
leads to a relation between $\eta_\kappa$, $\eta_u$ and the 
anisotropy exponent $z$
\begin{equation}
2\eta_\kappa + \eta_u = 4 - (D-1)/z
\end{equation}
which, together with the defining relation $z=2/(4-\eta_\kappa)$ reduces to
\begin{equation}
2\eta_u - (D-5)\eta_\kappa = 10-4D\;,
\label{eta_relation}
\end{equation}
and for $\eta_\kappa=0$ gives $z$ and $\eta_u$ in Eq.\ref{z_eta}.
This is supported by a detailed self-consistent one-loop
perturbative calculation of $g_y({\bf q})$ and by direct 
RG analysis.\cite{RTtubule}

For {\em phantom} membranes with $D=2$, $\eta_u=1$ and $z=1/2$, we
find:
\begin{eqnarray}
g_y({\bf q})&\propto&\left\{\begin{array}{lr}
q_y, & q_y>>\sqrt{q_\perp}\\
\sqrt{q_\perp}, &\;\;\; q_y<<\sqrt{q_\perp}\;.
\label{gy_limits}
\end{array} \right.
\end{eqnarray}
This leads to phonon rms fluctuations given by
\begin{equation}
\langle u({\bf x})^2\rangle=L_\perp^{1/4} S_u(L_y/L_\perp^{3/4})\;,
\label{uu_scaling_form}
\end{equation}
with universal scaling function having limiting form
\begin{eqnarray}
S_u(x)&\propto&\left\{\begin{array}{lr}
\mbox{constant}, & x\rightarrow\infty\\
x^{1/3}, & x\rightarrow 0\;.
\label{fu_limits}
\end{array} \right.
\end{eqnarray}
For roughly square membranes, $L_y\sim L_\perp=L$, so, as
$L\rightarrow\infty$, $L_y/L_\perp^{3/4}\rightarrow\infty$
this gives
\begin{equation}
\langle u({\bf x})^2\rangle=L_\perp^{1/4}\;,
\label{uu4}
\end{equation}
a result that is consistent with simulations by Bowick, et al..~\cite{BFT}

\subsubsection{Zero-modes and tubule shape correlation}

The tubule's shape is characterized by 
\begin{eqnarray}
R_G^2\equiv\langle|{\vec h}({\bf L}_\perp,y)- {\vec
h}(0_\perp,y)|^2\rangle\;,\;\;\;
h_{rms}^2\equiv\langle|{\vec h}({\bf x_\perp},L_y)- {\vec h}({\bf
x_\perp},0)|^2\rangle\;.\label{Rhdef}
\end{eqnarray}
As illustrated in Fig.\ref{tubuleShape} $R_G$ measures the 
radius of a typical cross-section of the tubule 
perpendicular to its extended (y-) axis, and $h_{rms}$ measures 
tubule end-to-end transverse fluctuations.

For a phantom tubule these are easily computed {\em exactly} using tubule free
energy, Eq.\ref{elFtubule}. One important subtlety is that one needs to take 
into account "zero modes" (Fourier modes
with ${\bf q_\perp}$ or $q_y=0$), that, because of anisotropic
scaling $q_\perp\sim q_y^2$ of the bulk modes can dominate tubule 
shape fluctuations.

For $R_G$ one finds
\begin{eqnarray}
R_G^2 &=& 2(d-D)\bigg[{k_B T\over L_y}\int_{L_\perp^{-1}}{d^{D-1}
q_\perp\over(2\pi)^{D-1}} {1\over t q_\perp^2}
(1 - e^{ i {\bf q}_\perp\cdot{\bf L}_\perp})\nonumber\\ 
&+& k_B T\int_{L_\perp^{-1},L_y^{-1}}{d^{D-1} q_\perp d q_y\over(2\pi)^D} 
{(1-e^{i {\bf q}_\perp\cdot {\bf L}_\perp})
\over t q_\perp^2 + \kappa({\bf q}) q_y^4}\bigg]\;,
\label{RgTubule}
\end{eqnarray}
with the first and second terms coming from the $q_y=0$ ``zero mode''
and the standard bulk contributions, respectively. From its definition, it is
clear that the ${\bf q}_\perp=0$ ''zero mode'' does not contribute to
$R_G$. Standard asymptotic analysis of above integrals gives
\begin{equation}
R_G(L_\perp,L_y)= L_\perp^{\nu}S_R(L_y/L_\perp^z)
\label{RG5}
\end{equation}
with, for phantom membranes, 
\begin{eqnarray}
\nu = {5-2D \over 4}\;,\;\;\;
z = {1\over2}\;,\label{nuz_phantom}
\end{eqnarray}
and the limiting form of the universal scaling function
$S_R(x)$ given by
\begin{equation}
S_R(x)\propto
\left\{ \begin{array}{lr}
1/\sqrt{x} & \mbox{for}\; x\rightarrow 0 \\
\mbox{constant}, & \mbox{for}\; x\rightarrow\infty\;.
\end{array} \right.
\label{fR}
\end{equation}
This gives
\begin{eqnarray}
R_G\propto L_\perp^{\nu}\propto L^{1/4},\;\; {\rm for}\; D=2\;,
\label{Rgdefine}
\end{eqnarray}
for the physically relevant case of a square membrane
$L_\perp\sim L_y\sim L\rightarrow\infty$, for which $L_y >>
L_\perp^z$, with bulk mode contribution dominating over the $q_y=0$ 
zero mode. This result is in excellent quantitative 
agreement with simulations of Bowick, et al.\cite{BFT} who found
$\nu=0.24\pm0.02$, in $D=2$. It would be interesting to test
the full anisotropic scaling prediction of Eqs.\ref{RG5},\ref{fR} 
by varying the
aspect ratio of the membrane in such simulations. For instance, for
fixed $L_\perp$ and increasing $L_y$ these predict 
{\em no} change in $R_G$. The same should hold if $L_y$ is {\em
decreased} at fixed $L_\perp$: $R_G$ should remain unchanged
until $L_y\sim\sqrt{L_\perp}$, at which point the tubule should begin
to get thinner (i.e. $R_G$ should decrease).

Equations \ref{RG5} and \ref{fR} also correctly recover the limit of
$L_y=\mbox{constant} << L_\perp^z\rightarrow\infty$, where the tubule
simply becomes a phantom, coiled up, $D-1$-dimensional polymeric network
of size $L_\perp$ embedded in $d-1$ dimensions, with the radius of
gyration $R_G(L_\perp)\sim L_\perp^{(3-D)/2}$. In the physical
dimensions ($D=2$ and $d=3$) this in particular gives a coiled up
ideal polymer of length $L_\perp$ with $R_G\sim L_\perp^{1/2}$, as
expected.

Similar analysis for the tubule roughness
$h_{rms}$ gives
\begin{eqnarray}
h_{rms}^2 &=& 2(d-D)\bigg[{k_B T\over L_\perp^{D-1}}\int_{L_y^{-1}}
{d q_y\over(2\pi)}
{1\over\kappa(q_y) q_y^4} (1-e^{i q_y L_y})
\;\nonumber\\
& + &k_B T\int_{L_\perp^{-1},L_y^{-1}}{d^{D-1} q_\perp d q_y\over(2\pi)^D}
{(1-e^{i q_y L_y})\over t q_\perp^2 +
\kappa({\bf q}) q_y^4}\bigg]\;.\label{hTubule}
\end{eqnarray}
In contrast to $R_G$, only the ${\bf q}_\perp=0$
zero mode (first term) and bulk modes contribute to $h_{rms}$, giving
\begin{equation}
h_{rms}(L_\perp,L_y)= L_y^{\zeta}S_h(L_y/L_\perp^z)
\label{h5}
\end{equation}
with, for phantom membranes, 
\begin{eqnarray}
\zeta = {5-2D \over 2}\;,\;\; z ={1\over2}\;.\label{z_phantom3}
\end{eqnarray}
and the asymptotics of $S_h(x)$ given by
\begin{equation}
S_h(x)\propto
\left\{ \begin{array}{lr}
\mbox{constant} & \mbox{for}\; x\rightarrow 0 \\
x^{3/2-\zeta}, & \mbox{for}\; x\rightarrow\infty
\end{array} \right.
\label{fh}
\end{equation}
Equations \ref{RG5} and \ref{h5} give information about the tubule
roughness for arbitrarily large size $L_\perp$ and $L_y$, and
arbitrary aspect ratio.  For the physically relevant case of a square
membrane $L_\perp\sim L_y\sim L\rightarrow\infty$, for which $ L_y >>
L_\perp^z$, in contrast to $R_G$ (where bulk modes dominates), 
${\bf q}_\perp=0$ zero mode dominates, leading to
\begin{equation}
h_{rms}\propto{L_y^{\zeta+(D-1)/2z}\over L_\perp^{(D-1)/2}}
\propto L^{\zeta+(D-1)(1-z)/2z}\;,
\label{hrmsL}
\end{equation}

Equations \ref{zeta}, \ref{z_phantom3} then give, for a $D=2$ phantom
tubule, $\zeta=1/2$, $z=1/2$
\begin{equation}
h_{rms}\sim {L_y^{3/2}\over L_\perp^{1/2}}\;,
\label{hrmsL1}
\end{equation}
and therefore predicts for a square phantom membrane wild transverse 
tubule undulations 
\begin{equation}
h_{rms}\sim L\;,
\label{hrmsL2}
\end{equation}
consistent with simulations\cite{BFT} that find 
$h_{rms}\sim L^\gamma$, with $\gamma=0.895\pm 0.06$.
As with $R_G$, it would be interesting to test the full scaling law
Eq.\ref{h5} by simulating non-square membranes, and testing for the
independent scaling of $h_{rms}$ with $L_y$ and $L_\perp$. Note that,
unlike $R_G$, according to Eq.\ref{hrmsL1}, $h_{rms}$ will show
immediate growth (reduction) when one increases (decreases) $L_y$ at
fixed $L_\perp$.

The above discussion also reveals that our earlier conclusions about
the lower critical dimension $D_{lc}$ for the existence of the tubule
are strongly dependent on how $L_\perp$ and $L_y$ go to infinity
relative to each other; i.e., on the membrane aspect ratio. The
earlier conclusion that $D_{lc}=3/2$ only strictly applies when the
bulk modes dominate the physics, which is the case for a very squat
membrane, with $ L_y\approx L_\perp^z$, in which case $L_y<<L_\perp$.
For the physically more relevant case of a square {\it phantom}
membrane, from the discussion above, we find that tubule phase is just
marginally stable with $D_{lc}=2^-$

Equations \ref{RG5} and \ref{h5} also correctly recover the limit of
$L_\perp^z=\mbox{constant}\ll L_y\rightarrow\infty$,
where the tubule simply becomes a polymer (ribbon) of thickness $R_G(L_\perp)$
of length $L_y$ embedded in $d-1$ dimensions.
These equations then correctly recover the polymer limit giving
\begin{equation}
h_{rms}\approx L_P (L_y/L_P)^{3/2}\;,
\end{equation}
with $L_\perp$-dependent persistent length $L_P(L_\perp)\propto
L_\perp^{D-1}$.  So, as
expected for a phantom tubule, if $L_\perp$ does not grow fast enough
(e.g. remains constant), while $L_y\rightarrow\infty$, the tubule
behaves as a linear polymer and crumples along its axis and the
distinction between the crumpled and tubule phases disappears.  

\subsection{Self-avoidance in the tubule phase}
\label{selfavoidance}
Self-avoidance is an important ingredient that must be included inside the
tubule phase. Detailed analysis of self-avoidance overturns 
arguments in Sec.\ref{tubuleAnomalous},
and leads to anomalous elasticity in the bending rigidity
modulus $\kappa({\bf q})$. Self-avoidance therefore also modifies the values of
other shape exponents, while leaving the scaling form of correlation
functions unchanged. 

In the y-tubule phase the self-avoidance interaction $F_{SA}$ from
Eq.\ref{Fc} reduces to
\begin{equation}
F_{SA}={v}\int\hspace{-0.1cm}dy\hspace{0.05cm} 
d^{D-1}x_\perp d^{D-1}x'_\perp
\hspace{0.05cm}\delta^{(d-1)}\big({\vec h}({\bf x_\perp},y)-
{\vec h}({\bf x'_\perp},y)\big)\;,
\label{TubuleSA}
\end{equation}
with $v=b/2\zeta_y$. 

\subsubsection{Flory theory}
\label{Flory}

The effects of self-avoidance can be estimated by
generalizing standard Flory arguments\cite{Flory_accurate} from polymer
physics\cite{Polymers} to the extended tubule geometry. The total
self-avoidance energy scales as
\begin{equation}
E_{SA}\propto V\rho^2\;,
\end{equation}
where $V\approx R_G^{d-1} L_y$
is the volume in the embedding space occupied by the tubule and
$\rho=M/V$ is the embedding space density of the tubule. Using the
fact that the tubule mass $M$ scales like $L_\perp^{D-1} L_y$, we see
that
\begin{equation}
E_{SA}\propto {L_y L_\perp^{2(D-1)}\over R_G^{d-1}}\;,\label{Esa}
\end{equation}

Using the radius of gyration $R_G\propto L_\perp^\nu$, and
considering, as required by the anisotropic scaling, a membrane with
$L_\perp\propto L_y^2$, we find that $E_{SA}\propto
L_y^{\lambda_{SA}}$ around the phantom fixed point, with
\begin{equation}
\lambda_{SA}=1+4(D-1)-(d-1)\nu\;,\label{lambda}
\end{equation}

Self-avoidance is relevant when $\lambda_{SA} > 0$, which, from the
above equation, happens for $\nu=\nu_{ph}=(5-2D)/4$ 
when the embedding dimension
\begin{equation}
d<d_{uc}^{SA}={6D-1\over5-2D}\;.\label{duc}
\end{equation}
For $D=2$-dimensional membranes, $d_{uc}^{SA} = 11$.  Thus, 
self-avoidance is strongly relevant for the tubule phase in $d=3$, in
contrast to the flat phase.

We can estimate the effect of the self-avoidance interactions on
$R_G$ in Flory theory, by balancing the estimate Eq.\ref{Esa}
for the self-avoidance energy with a similar estimate for the elastic
energy:
\begin{equation}
E_{elastic}=t\left({R_G\over L_\perp}\right)^2 L_\perp^{D-1} L_y\;.
\label{Eel}
\end{equation}
Equating $E_{elastic}$ with $E_{SA}$, we obtain a Flory estimate for
the radius of gyration $R_G$:
\begin{equation}
R_G(L_\perp)\propto L_\perp^{\nu_{F}}\;,\; {\nu_{F}}={D+1\over
d+1}\;,\label{SAnuDd}
\end{equation}
which should be contrasted with the Flory estimate of
$\nu_{F}^c=(D+2)/(d+2)$ for the {\em crumpled} phase. 
For the physical case $D=2$, Eq.\ref{SAnuDd} gives
\begin{equation}
R_G\propto L_\perp^{3/4}\;,\label{SAnu}
\end{equation}
a result that is known to be {\it exact} for the radius of gyration of
a $D=1$-polymer embedded in $d=2$-dimensions.\cite{2dPolymer} Since
the cross-section of the $D=2$-tubule, crudely speaking, traces out a
crumpled polymer embedded in two dimensions (see
Fig.\ref{tubuleShape}), it is intriguing to conjecture that ${\nu}=3/4$
is also the {\it exact} result for the scaling of the thickness of the
tubule.

\subsubsection{Renormalization group and scaling relations}
\label{RGscaling}
A new significant complexity that arises and is special to the tubule
phase (as compared to the crumpled phase) is the simultaneous presence
of local elastic and nonlocal (in the intrinsic space) self-avoidance 
nonlinearities. Above Flory mean-field analysis (that ignores elastic
nonlinearities) is nicely complemented by a renormalization group
approach that can handle this complexity of the full model $F=F_{el}+F_{SA}$, 
Eqs.\ref{elFtubule},\ref{TubuleSA}. Although, as we argued 
above, elastic nonlinearities are irrelevant in a phantom tubule,
in a physical self-avoiding tubule, they are indeed important and lead
to an anomalous bending rigidity elasticity.

The correct model, that incorporates the effects of both the
self-avoiding interaction and the anharmonic elasticity, is defined by
the full tubule free energy
\begin{eqnarray}
F&=&{1\over2}\int d^{D-1}x_\perp dy
\bigg[\kappa(\partial^2_y{\vec h})^2 + t(\partial^\perp_\alpha \vec h)^2
+ g_\perp(\partial^\perp_\alpha u)^2
+g_y\bigg(\partial_y u+{1\over2}(\partial_y{\vec h})^2\bigg)^2\bigg]\nonumber\\
&+&{v}\int\hspace{-0.1cm}dy\hspace{0.05cm} 
d^{D-1}x_\perp d^{D-1}x'_\perp
\hspace{0.05cm}\delta^{(d-1)}\big({\vec h}({\bf x_\perp},y)-
{\vec h}({\bf x'_\perp},y)\big)\;.
\label{SAtubuleRG}
\end{eqnarray}

To assess the role of elastic ($g_y$) and self-avoiding ($v$)
nonlinearities  we implement standard momentum-shell RG.\cite{Wilson}
We integrate out perturbatively in $g_y$ and $v$ 
short-scale fluctuations of modes $u({\bf q})$ and
$\vec{h}({\bf q})$ within a cylindrical shell 
$\Lambda e^{-l}<q_\perp<\Lambda$, $-\infty<q_y<\infty$, and 
anisotropically rescale lengths (${\bf x}_\perp$, $y$) and
fields (${\vec h}({\bf x}),u({\bf x})$), so as to restore the
ultraviolet cutoff to $\Lambda$:
\begin{eqnarray}
{\bf x}_\perp&=&e^{l}{\bf x}_\perp'\;,\;\;\; y=e^{z l} y' \;,\\
{\vec h}({\bf x})&=&e^{\nu l}{\vec h}'({\bf x}')\;,\;\;\;
u({\bf x})=e^{(2\nu-z) l}u'({\bf x}')\;,
\label{rescalings}
\end{eqnarray}
where we have chosen the convenient (but not necessary) rescaling of
the phonon field $u$ so as to preserve the form of the
rotation-invariant strain operator $\partial_y u+{1\over2}(\partial_y{\vec
h})^2$. Under this transformation the free energy returns back to
its form, Eq.\ref{SAtubuleRG}, but with 
effective length-scale ($l=\log L_\perp$) dependent coupling constants
determined by
\begin{eqnarray}
{d t\over d l}&=&[2\nu+z+D-3-f_t(v)]t\;,
\label{t_rr}\\
{d \kappa\over d l}&=&[2\nu-3z+D-1+f_\kappa(g_y,g_\perp)]\kappa\;,
\label{kappa_rr}\\
{d g_y\over d l}&=&[4\nu-3z+D-1-f_g(g_y)]g_y\;,
\label{gy_rr}\\
{d g_\perp\over d l}&=&[4\nu-z+D-3]g_\perp\;,
\label{gp_rr}\\
{d v\over d l}&=&[2D-2+z-(d-1)\nu-f_v(v)]v\;,
\label{v_rr}
\end{eqnarray}
where the various $f$-functions represent the graphical (i.e.,
perturbative) corrections.  Since the self-avoiding interaction only
involves $\vec{h}$, and (for convenience) 
the parameters in the $\vec{h}$ propagator
($t$ and $\kappa$) are going to be held fixed at $1$, the graphical
corrections coming from self-avoiding interaction alone depend only on
the strength $v$ of the self-avoiding interaction. Therefore, to {\em
all} orders in $v$, and leading order in $g_y$, $f_t(v)$ and $f_v(v)$
are only functions of $v$ and $f_\kappa(g_y,g_\perp)$ and $f_g(g_y)$
are only functions of $g_y$ and $g_\perp$.

It is important to note that $g_\perp$ suffers no graphical
corrections, i.e., Eq.\ref{gp_rr} is {\em exact}. This is enforced by
an exact symmetry
\begin{equation}
u({\bf x}_\perp,y)\rightarrow u({\bf x}_\perp,y)+\chi({\bf
x}_\perp)\;,
\label{u_symmetry}
\end{equation}
where $\chi({\bf x}_\perp)$ is an arbitrary function of ${\bf
x}_\perp$, under which the nonlinearities in $F$ are invariant.

We further note that there is an additional tubule ``gauge''-like
symmetry for $g_y=0$
\begin{equation}
\vec{h}({\bf x}_\perp,y)\rightarrow \vec{h}({\bf x}_\perp,y)+\vec{\phi}(y)\;,
\label{h_symmetry}
\end{equation}
under which the only remaining nonlinearity, the self-avoiding
interaction, being local in $y$, is invariant. This ``tubule gauge''
symmetry demands that $f_\kappa(g_y=0,g_\perp)=0$, which implies that
if $g_y=0$, there is no divergent renormalization of $\kappa$, {\em
exactly}, i.e., the self-avoiding interaction {\em alone} cannot
renormalize $\kappa$. This {\em non}-renormalization of $\kappa$ by
the self-avoiding interaction, in a truncated (unphysical) membrane
model with $g_y=0$, has been verified to all orders in a
perturbative renormalization group calculation.\cite{BG}

To see that $\nu$ and $z$ obtained as fixed point solutions of
Eqs.\ref{t_rr}-\ref{v_rr} have the same physical significance as the
$\nu$ and $z$ defined in the scaling expressions Eqs.\ref{hrms} 
for the radius of gyration $R_G$ and tubule wigglyness
$h_{rms}$, we use RG transformation to relate
these quantities in the unrenormalized system to those in the
renormalized one. This gives, for instance, for the radius of gyration
\begin{eqnarray}
R_G(L_\perp,L_y;t,\kappa,\ldots)
&=&e^{\nu l} R_G(e^{-l}L_\perp,e^{-z l} L_y;t(l),\kappa(l),\ldots)\;,
\nonumber\\
\label{matching1}
\end{eqnarray}
Choosing $l=l_*=\log L_\perp$ this becomes:
\begin{equation}
R_G(L_\perp,L_y;t,\kappa,\ldots)=
L_\perp^\nu R_G(1,L_y/L_\perp^z;t(l_*),\kappa(l_*),\ldots)\;.
\label{matching2}
\end{equation}
This relation holds for {\it any} choice of the (after all, arbitrary)
rescaling exponents $\nu$ and $z$. However, {\it if} we make the
special choice such that Eqs.\ref{t_rr}-\ref{v_rr} lead to fixed
points, $t(l_*)$,
$\kappa(l_*),\ldots$ in Eq.\ref{matching2} go to {\it constants},
independent of $l_*$ (and hence $L_\perp$), as $L_\perp$ and hence
$l_*$, go to infinity. Thus, in this limit, we obtain from
Eq.\ref{matching2}
\begin{equation}
R_G(L_\perp,L_y;t,\kappa,\ldots)=
L_\perp^\nu R_G(1,L_y/L_\perp^z;t_*,\kappa_*,\ldots)\;,
\label{matching3}
\end{equation}
where $t_*, \kappa_*, \ldots$ are the fixed point values of coupling
constants. This result clearly agrees with the scaling forms for
$R_G$, Eq.\ref{hrms} (with analogous derivation for $h_{rms}$) with the
scaling function given by $S_R(x)\equiv R_G(1,x;t_*,\kappa_*,g_y^*,v^*)$.

The recursion relations Eqs.\ref{t_rr}-\ref{v_rr} reproduce all of
our earlier phantom membrane results (when $v=0$, leading to
$f_\kappa=0$), as well as the upper-critical embedding
dimension $d_{uc}^{SA}=(6D-1)/(5-2D)$ for self-avoidance predicted by 
Flory theory, Eq.\ref{lambda}, {\em and} the upper critical {\it intrinsic}
dimension $D_{uc}=5/2$ for anomalous elasticity for phantom
membranes (determined by eigenvalues of dimensionless couplings
corresponding to $v$ and $g_y$). They also reproduce all of the Flory
theory exponents under the approximation that graphical corrections to
$t$ and $v$ are the same, i.e., that $f_t(v_*) = f_v(v_*)$.

To analyze the renormalization of $\kappa$ in a self-avoiding
tubule, we focus once again on the nonlocal interaction
the non-local interaction $F_{anh}$, Eq.\ref{Fanh}, mediated
by integrated out phonons, with a kernel
\begin{equation}
V_h({\bf q})={g_y g_\perp q_\perp^2\over g_y q_y^2+g_\perp q_\perp^2}\;,
\label{vertexV_h2}
\end{equation}
whose long-scale scaling determines renormalization of $\kappa$. 
If $g_y({\bf q}) q_y^2 >> g_\perp({\bf q}) q_\perp^2$
(as we saw for a phantom tubule) then at long scales $V_h({\bf
q})\approx g_\perp q_\perp^2/q_y^2\sim q_y^2$
in the relevant limit of $q_\perp\sim q_y^2$. Simple
power counting around the Gaussian fixed point then shows that this
elastic nonlinearity is irrelevant for $D>D_{uc}=3/2$, and therefore
$f_\kappa^*=\eta_\kappa=0$ for a physical $D=2$-dimensional tubule, 
as we argued in Sec.\ref{tubuleF}.

On the other hand, if the scaling is such that $g_\perp({\bf q})
q_\perp^2$ dominates over $g_y({\bf q}) q_y^2$, then $V_h({\bf
q})\approx g_y$, i.e. a constant at long length scales. Simple
power-counting then shows that this coupling is relevant for
$D<D_{uc}=5/2$ and the bending rigidity modulus of a $D=2$-dimensional
tubule {\it is} anomalous in this case.

In the RG language, the relevance of $V_h$ is decided by the sign of the
renormalization group flow eigenvalue of $g_\perp(l)$ in
Eq.\ref{gp_rr}
\begin{equation}
\lambda_{g_\perp}=4\nu-z+D-3\;,
\label{lambda_gp}
\end{equation}
which is exactly determined by the fixed-point values of $\nu$ and $z$, since
$g_\perp$ suffers no graphical renormalization. 

As discussed in previous sections, for a phantom tubule (or for $d>d_{uc}$)
$\nu=(5-2D)/4$ and $z=1/2$. For $d$ below but close to 
$d_{uc}^{SA}=(6D-1)/(5-2D)$ ($=11$
for $D=2$), these values are modified by the self-avoiding
interaction, but only by order $\epsilon\equiv
d-d_{uc}^{SA}$. 
Hence a $D=2$-dimensional tubule, embedded in $d$ dimensions close to
$d_{uc}^{SA}=11$, $\lambda_{g_\perp}\approx-1/2$ and $g_\perp(l)$ flows
according to
\begin{equation}
{d g_\perp\over d l}=[-{1\over 2} + O(\epsilon)]g_\perp\;,
\label{gp_rr11}
\end{equation}
i.e. $g_\perp$ is {\em irrelevant} near $d=11$ (for $\epsilon\ll 1$),
$V_h({\bf q})\sim g_\perp q_\perp^2/q_y^2\sim q_y^{2-O(\epsilon)}$ is
irrelevant to the {\it bend} elasticity for a 
$D=2$-dimensional tubule embedded in high dimensions near
$d_{uc}^{SA}=11$, and, hence, $f_\kappa=\eta_\kappa$ in Eq.\ref{kappa_rr}
vanishes as the fixed point. Therefore, in these high embedding dimensions 
the full model of a self-avoiding tubule 
reduces to the {\em linear} elastic truncated model with 
self-avoiding interaction as the only nonlinearity, that 
can be nicely studied by standard expansion in 
$\epsilon=d_{uc}^{SA}-d$.\cite{BG} As we
discussed above, the ``tubule gauge'' symmetry guarantees that in this
case the self-avoiding interaction alone cannot renormalize $\kappa$,
i.e., $f_\kappa=\eta_\kappa=0$ for $d$ near $d_{uc}^{SA}$. This
together with Eq.\ref{kappa_rr} at its fixed point leads to an {\it
  exact} exponent relation 
\begin{equation}
z={1\over 3}(2\nu+D-1)\;,
\label{z_nu_phantom}
\end{equation}
valid for a finite {\it range} $d_*<d<d_{uc}^{SA}$ of
embedding dimensions, and for phantom tubules in any embedding
dimension. Where valid, it therefore reduces the tubule problem to a 
single independent shape exponent.

However, this simple scenario, and, in particular, the scaling
relation Eq.\ref{z_nu_phantom}, is {\it guaranteed} to break down as
$d$ is reduced. The reason for this is that, as $d$ decreases, $\nu$
increases, and eventually becomes so large that the eigenvalue
$\lambda_{g_\perp}$ of $g_\perp$ changes sign and becomes positive. As
discussed earlier, once this happens, the nonlinear vertex
Eq.\ref{vertexV_h2} becomes relevant, and $\kappa$ acquires a
divergent renormalization, i.e., $f_\kappa\neq 0$, and bend tubule
elasticity becomes anomalous. 

Now, it is easy to show, using Eq.\ref{z_nu_phantom} and a rigorous 
lower bound $\nu > (D-1)/(d-1)$ inside $\lambda_{g_\perp}$,
that it {\it must} become positive for $d>d_*^{lb}(D)$ with
\begin{eqnarray}
d_*^{lb}(D)&=&{4D-1\over 4-D}=7/2\;,\;\;{\rm for}\;D=2.\label{d_star2}
\end{eqnarray}
Hence, for the case of interest $D=2$ critical dimension $d_*$ is bounded 
by $7/2$ from below.  In fact, Flory and $\epsilon=11-d$ estimates indicate
that $d_*(D=2)\approx 6$.\cite{RTtubule}

\begin{figure}[bth]
\centering
\setlength{\unitlength}{1mm}
\begin{picture}(150,70)(0,0)
\put(-5,-72){\begin{picture}(150,70)(0,0)
\includegraphics{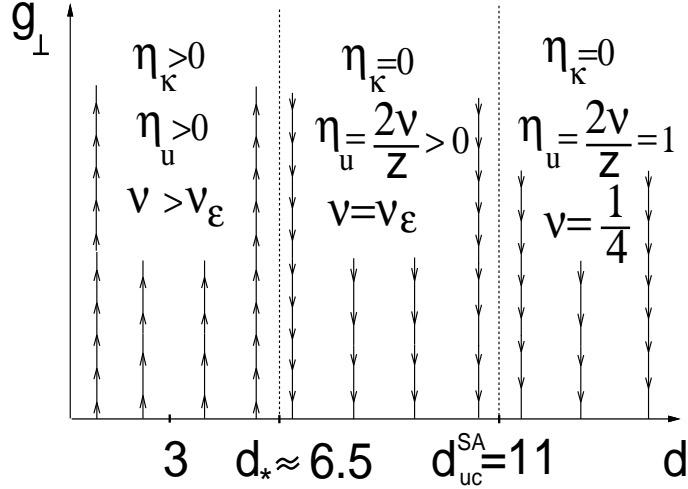}
\end{picture}}
\end{picture}
\caption{Schematic illustration of change in
relevance of $g_\perp(l)$ at $d_*$. For embedding
dimensions below $d_*$ (which includes the physical case of $d=3$),
$g_\perp(l)$ becomes relevant and (among other phenomena) leads to 
anomalous bending elasticity with 
$\kappa({\bf q})\sim q_y^{-\eta_\kappa}$, that diverges at long
length scales.}
\label{gperp_d}
\end{figure}

\begin{figure}[bth]
\centering
\setlength{\unitlength}{1mm}
\begin{picture}(150,80)(0,0)
\put(-10,-85){\begin{picture}(150,70)(0,0)
\includegraphics{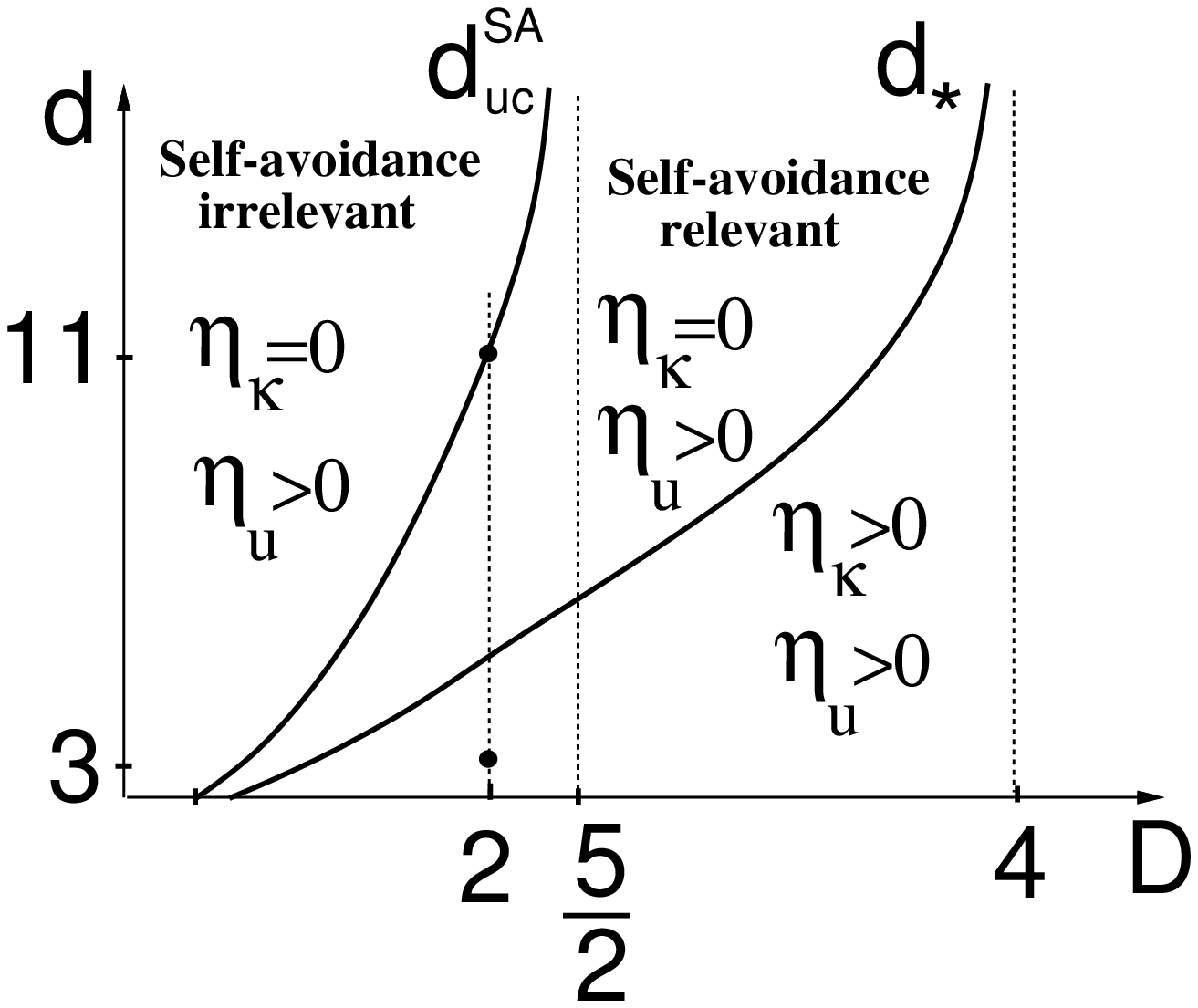}
\end{picture}}
\end{picture}
\caption{Schematic of the tubule ``phase'' diagram in the embedding
$d$ vs intrinsic $D$ dimensions. Self-avoiding interaction becomes
relevant for $d<d_{uc}^{SA}(D)=(6D-1)/(5-2D)$, ($=11$, for
$D=2$). Below the $d_*(D)$ curve (for which the lower bound is
$d_*^{lb}(D)=(4D-1)/(4-D)$) the anharmonic elasticity becomes
relevant, leading to anomalous elasticity with a divergent bending
rigidity.}
\label{d_vs_D}
\end{figure}

We therefore conclude that in a physical
$D=2$-dimensional self-avoiding tubule, embedded in $d=3 < d_*\approx
6$, anharmonic elasticity $F_{anh}$ {\em is} important at long scales and
leads to anomalous and divergent bending rigidity $\kappa({\bf q})$
and $g_y({\bf q})$
\begin{eqnarray}
g_y({\bf q})=q_y^{\eta_u} S_g(q_y/q_\perp^z)\;,\;\;
\kappa({\bf q})=q_y^{-\eta_\kappa}
S_\kappa(q_y/q_\perp^z)\;,
\label{kappaScaleForm}
\end{eqnarray}
with 
\begin{eqnarray}
z\eta_\kappa=f_\kappa(g_y^*, g_\perp^*)\;,\;\;
z\eta_u=f_g(g_y^*)\;.
\label{eta_u}
\end{eqnarray}
and asymptotic forms of scaling functions
\begin{eqnarray}
S_g(x\rightarrow 0)\rightarrow x^{-\eta_u}\;,\;\;
S_\kappa(x\rightarrow 0)\rightarrow x^{\eta_\kappa}\;.
\label{S_kappa}
\end{eqnarray}
Another consequence is the breakdown of the exponent relation, 
Eq.\ref{z_nu_phantom}, that is replaced by {\em two} 
exact relations holding between
{\em four} independent exponents $z$, $\nu$, $\eta_\kappa$, and
$\eta_u$ 
\begin{eqnarray}
z={1\over 3-\eta_\kappa}(2\nu+D-1)\;,\;\;\;
z={1\over 3+\eta_u}(4\nu+D-1)\;,
\label{z_nu_new2}
\end{eqnarray}
which automatically contain the rotational symmetry Ward identity
\begin{equation}
2\eta_\kappa+\eta_u=3-(D-1)/z\;,
\label{rot_symm}
\end{equation}
formally arising from the requirement that
graphical corrections do not change the form of the rotationally
invariant strain operator $\partial_y u+{1\over2}(\partial_y{\vec h})^2$.
The existence of a nontrivial $d_*>3$ and its consequences are summarized
by Figs.\ref{gperp_d},\ref{d_vs_D}.

\begin{figure}[bth]
\centering
\setlength{\unitlength}{1mm}
\begin{picture}(150,57)(0,0)
\put(-5,-105){\begin{picture}(150,57)(0,0)
\includegraphics{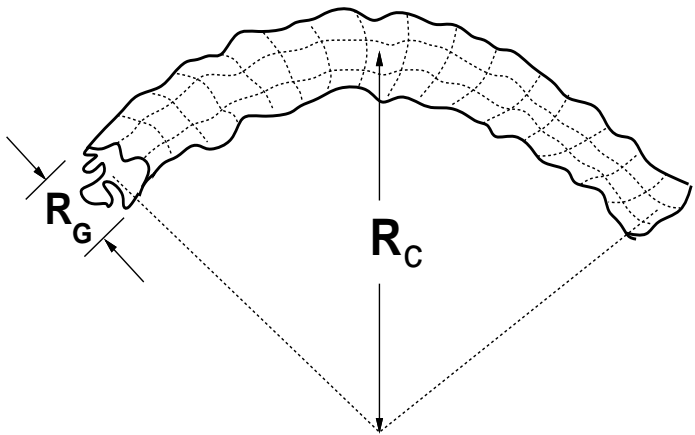}
\end{picture}}
\end{picture}
\caption{Illustration of the physical mechanism for the enhancement of
the bending rigidity $\kappa$ by the shear $g_y$ elasticity. To bend a
polymerized tubule of thickness $R_G$ into an arc of radius $R_c$
requires $R_G/R_c$ fraction of bond stretching and therefore costs
elastic shear energy, which when interpreted as bending energy leads
to a length-scale dependent renormalization of the bending rigidity
$\kappa$ and to the Ward identity Eq.\ref{expRelation}.}
\label{bend_tubule}
\end{figure}

The physics behind above somewhat formal derivation of exponent
relations (Ward identities) can be further
exposed through a simple physical shell argument.  As can be seen from 
Fig.\ref{bend_tubule}, bending of a tubule of radius $R_G$ into an
arc of radius $R_c$ induces an in-plane strain energy 
density $g_y(L_y,L_\perp)(R_G(L_y)/R_c)^2$.
Interpreting this additional energy as an effective bending energy
density $\kappa_y(L_\perp , L_y)/R_c^2$, leads to the {\it effective}
bending modulus $\kappa_y(L_\perp ,L_y)$,
\begin{equation}
\kappa_y(L_\perp,L_y)\sim
g_y(L_\perp,L_y)R_G(L_\perp,L_y)^2\;.\label{kappaInduced}
\end{equation}
This leads to a relation between the scaling exponents
\begin{equation}
2\nu=z(\eta_\kappa+\eta_u)\;.\label{expRelation}
\end{equation}
that is contained in Eqs.\ref{z_nu_new2} obtained 
through RG analysis above.

We note, finally, that all of the exponents must show a jump
discontinuity at $d_*$. Therefore,
unfortunately, an extrapolation from $\epsilon=11-d$--expansion in a
truncated model with linear elasticity\cite{BG} down to the physical
dimension of $d=3$ (which is rigorously below $d_*$) gives little information
about the properties of a real tubule.
%
%

The computations for a physical tubule must be performed for $d<d_*$,
where both the self-avoidance and the elastic nonlinearities are
both relevant and must be handled simultaneously. As we discussed above,
for $d<d_*$, the eigenvalue $\lambda_{g_\perp}>0$, leading to the flow
of $g_\perp(l)$ to infinity, which in turn leads to $V_h({\bf q})=
g_y$. Physically this regime of $g_\perp\rightarrow\infty$ corresponds
to freezing out the phonons $u$, i.e. setting $u=0$ in the free energy
$F[\vec{h},u]$, Eq.\ref{SAtubuleRG},
with the resulting effective free energy functional for a
physical self-avoiding tubule given by
\begin{eqnarray}
F&=&{1\over2}\int d^{D-1}x_\perp dy
\bigg[\kappa(\partial^2_y{\vec h})^2 + t(\partial^\perp_\alpha \vec h)^2
+ {1\over4}g_y(\partial_y{\vec h})^4\bigg]\nonumber\\
&+&{v}\int\hspace{-0.1cm}dy\hspace{0.05cm} 
d^{D-1}x_\perp d^{D-1}x'_\perp
\hspace{0.05cm}\delta^{(d-1)}\big({\vec h}({\bf x_\perp},y)-
{\vec h}({\bf x'_\perp},y)\big)\;,
\label{SAtubuleRGeff}
\end{eqnarray}
Since one must perturb in $g_y$ around a nontrivial, {\it
strong} coupling fixed point with $v^*=O(1)$, unfortunately, no
controlled analysis of above model exists todate and remains a
challenging open problem. Nevertheless, above RG analysis combined
with Flory estimates and exact exponent relations provides
considerable information about shape fluctuations and elasticity of 
a polymerized tubule.

\subsection{Phase transitions}
\label{transition_section}

Now that we have solidly established the properties of the three
phases of anisotropic polymerized membranes, we turn to analysis of
phase transitions between them. As discussed in the Introduction, 
a direct crumpled-to-flat transition 
is highly nongeneric for anisotropic membranes, as it has to be finely
tuned to the tetracritical point illustrated in
Fig.\ref{PhaseDiagram_tubuleF}. If so tuned this transition will be in the
same universality class as for isotropic membranes (where it is
generic).\cite{PKN,LRscsa} 

Here we will focus on the new crumpled-to-tubule and
tubule-to-flat transitions, which are generic for membranes with any amount of
anisotropy. As for the tubule phase itself and the crumpled-to-flat
transition,  a complete analysis of these transitions 
must include both elastic and self-avoiding nonlinearities, a highly
nontrivial open problem. Below we will instead present an incomplete
solution. First we will present an RG analysis of a phantom
(non-selfavoiding) membrane, focusing on the much simpler
crumpled-to-tubule transition. We will then complement this study with a
scaling theory and Flory approximation of the crumpled-to-tubule and
tubule-to-flat transitions in a more realistic self-avoiding membrane.

\subsubsection{Renormalization group analysis of crumpled-to-tubule
transition}
\label{crumpled-to-tubuleRG}
The crumpled-to-(y-)tubule (CT) transition takes place when $t_y\rightarrow
0$, while $t_\perp$ remains positive. This implies that the CT
critical point is characterized by highly anisotropic 
scaling $q_\perp\propto q_y^2$. It leads to a considerable
simplification of the full free energy defined in Eq.\ref{Fc} down to
%
\begin{eqnarray}
F[{\vec r}({\bf x})]= {1\over2}\int d^{D-1}x_\perp dy
\bigg[\kappa_y\left(\partial_y^2\vec{r}\right)^2
+ t_\perp\left(\partial^\perp_\alpha\vec{r}\right)^2
+ t_y\left(\partial_y\vec{r}\right)^2 +
{u_{yy}\over2}\left(\partial_y\vec{r}\cdot\partial_y\vec{r}\right)^2\bigg]\;.
\nonumber\\
\label{Fcii}
\end{eqnarray}
The standard $O(d)$ $\phi^4$ model form of $F$ facilitates 
systematic analysis using conventional RG methods.\cite{Wilson} 
Because of the strong {\em scaling} anisotropy of the quadratic pieces 
of the free energy, it is convenient to rescale lengths 
${\bf x}_\perp$ and $y$ anisotropically:
\begin{eqnarray}
{\bf x}_\perp={\bf x}_\perp'e^{l}\;,\;\;\;y=y' e^{z l}\;,
\end{eqnarray}
and rescale the ``fields'' ${\vec r}({\bf x})$ according to 
\begin{equation}
{\vec r}({\bf x})=e^{\chi l}{\vec r}'({\bf x}')\;.
\end{equation}
Under this transformation
\begin{eqnarray}
\kappa_y(l)=\kappa_y e^{(D-1-3z+2\chi)l}\;,\;\;\;
t_\perp(l)=t_\perp e^{(D-3+z+2\chi)l}\;.\label{t_scale}
\end{eqnarray}
Requiring that both $\kappa_y$ and $t_\perp$ remain fixed under this
rescaling (zeroth order RG transformation) fixes the ``anisotropy''
exponent $z$ and the ``roughness'' exponent $\chi$:
\begin{eqnarray}
z={1\over2}\;,\;\;\;
\chi=(5/2-D)/2\;.\label{chi_quad}
\end{eqnarray}
Although this choice keeps the quadratic in $\vec{r}$ part of $F$,
Eq.\ref{Fcii}, unchanged, it {\em does} change the quartic piece:
\begin{eqnarray}
u_{yy}(l)= u_{yy}e^{(D-1-3z+4\chi)l}=u_{yy}e^{(5/2-D)l}\;,
\label{uyy_scale}\;
\end{eqnarray}
and shows that below the upper critical
dimension $D_{uc}=5/2$, the Gaussian critical point is unstable to 
elastic nonlinearities, that become comparable to harmonic elastic 
energies on scales longer than the characteristic
length scale $L_\perp^{nl}$
\begin{equation}
L_\perp^{nl}=\left(\kappa_y\over u_{yy}\right)^{1/(5/2-D)}\;.
\label{Lnl}
\end{equation}
To describe the new behavior that prevails on even {\em longer} length
scales requires a full-blown RG treatment.

Such an analysis\cite{RTtubule} leads to corrections to the simple
rescaling of $\kappa_y$, $t_\perp$, and $t_y$ (due to
$u_{yy}$ non-linearity), characterized by ``anomalous'' exponents
$\eta_\kappa$, $\eta_t$, and $\delta\theta$:
\begin{eqnarray}
\kappa_y(l)&=& \kappa_y e^{(D-1-3z+z\eta_\kappa +2\chi) l}\;,\\
\label{kappa}
t_\perp(l)&=& t_\perp e^{(D-3+z+\eta_t+2\chi) l}\;,\\
\label{tperp}
t_y(l)&=& t_y e^{(D-1-z-\delta\theta+2\chi) l}\equiv t_y e^{\lambda_t l}\;,\\
\label{ty}
\nonumber
\end{eqnarray}
that give
\begin{eqnarray}
z={2-\eta_t\over4-\eta_\kappa}\;,\;\;\;
\chi= {10-4D+\eta_\kappa(D-3+\eta_t)-3\eta_t\over8-2\eta_\kappa}\;,
\label{chi}
\nonumber
\end{eqnarray}
valid at the new nontrivial CT critical point. 

Once the values of $\eta_t$, $\eta_\kappa$ and $\chi$ at the critical
point are determined, the renormalization group gives a relation
between correlation functions at or near criticality (small $t_y$) and
at small wavectors (functions that are difficult to compute, because
direct perturbation theory is divergent) to the same correlation
functions away from criticality and at large wavevectors (functions
that can be accurately computed using perturbation theory). For
example the behavior of the correlation lengths near the transition
can be deduced in this way:
\begin{eqnarray}
\xi_\perp(t_y)= e^{l}\xi_\perp(t_y e^{\lambda_t l})\;,\;\;\;
\xi_y(t_y)= e^{z l}\xi_y(t_y e^{\lambda_t l})\;,
\label{xiy}
\end{eqnarray}
assuming that a critical fixed point exists and
all other coupling constants have well-defined values at this point. 
Taking $t_y e^{\lambda_t l}\approx 1$ then gives
\begin{eqnarray}
\xi_\perp(t_y)\approx a\; t_y^{-\nu_\perp}\;,\;\;\;
\xi_y(t_y)\approx a\; t_y^{-\nu_y}\;, 
\label{xiyii}
\end{eqnarray}
where $a\approx\xi(1)$ is the microscopic cutoff and,
\begin{eqnarray}
\nu_\perp&=&{1\over\lambda_t}
={4-\eta_\kappa\over 2(2-\eta_t-2\delta\theta)-
\eta_\kappa(2-\eta_t-\delta\theta)}\;,\;\;\;
\nu_y={z\nu_\perp}\;
\label{nuy}
\end{eqnarray}
are correlation length exponents perpendicular and along the tubule axis.

The anomalous exponents can be computed by
integrating out short-scale degrees of freedom perturbatively in
$u_{yy}$. Standard analysis shows that indeed there is a nontrivial
critical point (at a finite value of $u_{yy}^*=O(\epsilon)$, 
$\epsilon\equiv 5/2-D)$), at which, to all orders $\eta_t=0$, and 
to one-loop order, for a physical membrane ($D=2$, $d=3$, i.e., $\epsilon=1/2$)
\begin{eqnarray}
\eta_\kappa=0,\;\; z=1/2, \;\;
\chi=1/4,\;\;
\nu_\perp\approx 1.227\;,\;\;\nu_y&\approx& 0.614\;.
\label{nuy2D}
\end{eqnarray}
It is interesting to note that, in contrast to the treatment of 
crumpled-to-flat transition in isotropic membranes\cite{PKN,LRscsa}, 
where the critical point was only
stable for an unphysically large value of the embedding dimension
$d>219$, the critical point characterizing the crumpled-to-tubule
transition discussed here is stable for all $d$. Furthermore, the
relatively small value of $\epsilon=1/2$ (in contrast to for example
$\epsilon=2$ for flat phase and crumpling transition) gives some 
confidence that above critical exponents for the CT transition in a 
phantom membrane might even be quantitatively trustworthy.

\subsubsection{Scaling theory of crumpled-to-tubule and tubule-to-flat
transitions}
\label{scaling_theory}
Above RG analysis for the CT transition in phantom membranes, can be
nicely complemented by a scaling theory combined with Flory estimates,
that can incorporate both the elastic and self-avoiding
nonlinearities, as we now describe.

Standard scaling arguments, supported by RG
analysis and matching calculations (see e.g., Sec.\ref{tubuleF})
suggest that near the crumpled-to-tubule transition, for a square membrane of
internal size $L$, membrane extensions $R_y$ and $R_G$ along and orthogonal to
the developing tubule axis should exhibit scaling form:
\begin{eqnarray}
R_{G, y}&=& L^{\nu_{ct}^{G,y}} f_{G,y} (t_y L^{\phi})\;,\nonumber\\
&\propto&
\left\{ \begin{array}{lr}
t_y^{\gamma_{+}^{G,y}} L^{\nu_c}, & t_y > 0,  L >> \xi_{ct} \\
L^{\nu_{ct}^{G,y}} , & L << \xi_{ct} \\
|t_y|^{\gamma_{-}^{G,y}} L^{\nu_{t}^{G,y}}, & t_y < 0, L >> \xi_{ct}
\label{Rtrans}
\end{array} \right.
\end{eqnarray}
where subscripts $t$, $c$ and $ct$ refer to tubule, crumpled and
tubule-to-crumpled transition, respectively, and $\xi_{ct}\propto
|t_y|^{-1/\phi}$ is a correlation length for the crumpled-to-tubule
transition, $t_y=(T-T_{ct})/T_{ct}$, $T_{ct}$ is the
crumpled-to-tubule transition temperature, with $t_y>0$ corresponding to
the crumpled phase. Consistency demands that exponents $\gamma_{+/-}^{G,y}$ 
are given by
\begin{eqnarray}
\gamma_{+}^{G,y}={\nu_c-\nu_{ct}^{G,y}\over\phi}\;,\;\;
\gamma_{-}^{G,y}={\nu_t^{G,y}-\nu_{ct}^{G,y}\over\phi}\;.
\label{scaling_below}
\end{eqnarray}

The asymptotic forms in Eq.\ref{Rtrans} are dictated by general
defining properties of the phases and the CT transition. For example,
scaling of both $R_{y}$ and $R_G$ like $L^{\nu_c}$, with the same 
exponent $\nu_c$ in the crumpled phase is rooted in the isotropy of
that phase. In contrast, extended and highly anisotropic nature of 
the tubule phase dictates that $\nu_t^G\neq\nu_t^y=1$. 
The anisotropy is, however, manifested even in the crumpled
phase through the different temperature-dependent amplitudes of $R_G$ and
$R_y$, with the aspect ratio $R_y/R_G$ actually {\it
  diverging} as $T\rightarrow T^+_{ct}$, and membrane begins to 
extend into a tubule configuration.
The former of these vanishes as $t_y\rightarrow 0^+$ (since the
radius of gyration in the tubule phase is much less than that in the
crumpled phase, since $\nu_t < \nu_c$), which implies $\gamma_+^G >
0$, while the latter diverges as $t_y\rightarrow 0^+$, since the
tubule ultimately extends in that direction, which implies $\gamma_+^y
<0$. 

We will now show how these general expectations are born out by
the Flory theory. Following analysis very similar to that done in 
Sec.\ref{Flory},
but keeping track of temperature-dependent order
parameter $\zeta_y$, we find that the Flory approximation to the 
free energy density in Eq.\ref{Fcii}, supplemented with 
self-avoidance is given by
\begin{equation}
\hspace{-.5cm}f_{Fl}[R_G,\zeta_y]=t_y \zeta_y^2 + u_{yy} \zeta_y^4 +
t_\perp\left({R_G\over L_\perp}\right)^2 +
v {L_\perp^{D-1}\over \zeta_y
R_G^{d-1}}\;.
\label{Ecritub}
\end{equation}
Minimizing this over $R_G$, gives
\begin{equation}
R_G\approx
L_\perp^{\nu_t} \bigg({ v \over t_\perp \zeta_y}\bigg)^{1/(d + 1)}\;,
\label{RGcrit}
\end{equation}
with the tubule exponent $\nu_t= {D+1 \over d+1}$ found earlier, 
but now with additional temperature and $L_\perp$ dependence of $R_G$
through $\zeta_y(t_y,L_\perp)$ that interpolates between tubule,
crumpled and critical behavior.  Inserting this expression for $R_G$ into
Eq.\ref{Ecritub}, gives
\begin{equation}
\hspace{-0.3cm}f_{Fl}[\zeta_y]=t_y \zeta_y^2 + 
u_{yy} \zeta_y^4 + t_\perp^{d-1\over d+1}
\left(v\over\zeta_y\right)^{2\over d+1}L_{\perp}^{-{2(d-D)\over
d+1}}\;.
\label{Ecritub2}
\end{equation}

Minimizing $f_{Fl}[\zeta_y]$ in the crumpled phase ($t_y>0$)
gives
\begin{equation}
\zeta_y\approx \left({v^2 t_\perp^{d-1}\over
t_y^{d+1}}\right)^{1\over2(d+2)} L_\perp^{-{d-D\over d+2}}\;,
\label{zeta_y2}
\end{equation}
that, as expected vanishes in the thermodynamic limit. For a square
($L\times L$) membrane this then gives 
for $R_y=\zeta_y L_y$ and $R_G$ (using Eq.\ref{RGcrit})
\begin{equation}
R_y\propto t_y^{-{d+1\over2(d+2)}} L_\perp^{{D+2\over d+2}}\;,\;\;
R_G\propto t_y^{{1\over2(d+2)}} L_\perp^{{D+2\over d+2}}\;,
\end{equation}
which, after comparing with the general form, Eq.\ref{Rtrans}, gives 
\begin{eqnarray}
\nu_c={D+2\over d+2}\;,\;\;
\gamma_+^y=-{d+1\over 2(d+2)}\;,\;\;
\gamma_+^G={1\over2(d+2)}\;,\label{nu_gamma}
\end{eqnarray}
$\nu_c$ reassuringly agrees with the well-known Flory result 
for the radius of gyration exponent $\nu_c$ for a $D$-dimensional 
manifold, embedded in $d$ dimensions,\cite{KKN,KNsa,KKNsa,ALsa} and 
$\gamma_+^{y,G}$ special to crumpled {\em anisotropic} membranes. 

As anticipated earlier, $\gamma_+^y\neq\gamma_+^G$ implies
that intrinsically anisotropic membrane are qualitatively distinct
from isotropic ones even in the crumpled phase, as they exhibit
a ratio of major to minor moment of inertia eigenvalues 
(related to $R_G/R_y$) that diverges as CT transition is 
approached.

Now for the tubule phase, characterized by $t_y<0$ and a finite order
parameter $\zeta_y>0$, last term in $f_{Fl}$, Eq.\ref{Ecritub2} 
is clearly negligible for $L_\perp > \xi_{cr}$
and simple minimization gives $\zeta_y \approx\sqrt{|t_y|/u_{yy}}$,
which, when then inserting into $R_{y,G}$ and 
comparing with the general scaling forms gives for a
square membrane
\begin{eqnarray}
\nu_t^y=1\;,\;\;
\gamma_{-}^y={1\over2}\;,\;\;
\nu_t^G={D+1\over d+1}\;,\;\;
\gamma_-^G=- {1 \over 2 ( d + 1 ) }\;.\label{nu_gammaT}
\end{eqnarray}

Finally, right at the crumpled-to-tubule transition, $t_y = 0$,
minimization of $f_{Fl}[\zeta_y]$ gives
\begin{equation}
\zeta_y \propto L_\perp^{- {(d - D) \over 3 + 2d} }
\label{zetaycrit}
\end{equation}
which, when inserted in $R_{y,G}$ gives the advertised critical scaling forms
with exponents
\begin{equation}
\nu_{ct}^y = { D + d + 3 \over 2d + 3}\;,\;\;
\nu_{ct}^G= { 2D + 3 \over 2d + 3}\;,\;\;
\phi = {2 ( d - D) \over 2d + 3}\;.
\label{nu_ct}
\end{equation}
that are reassuringly consistent with our independent calculations of
exponents $\gamma_{+,-}^{G,y}$, $\nu_c$, $\nu_t^{G,y}$, and
$\nu_{ct}^{G,y}$ using exact exponent relations above.
For the physical case of a two dimensional membrane embedded
in a three dimensions, ($D=2,d=3$)
\begin{eqnarray}
\nu_c&=&4/5\;,\;\;
\nu_{ct}^G=7/9\;,\;\;
\nu_{ct}^y=8/9\;,\;\;
\nu_t=3/4\;,\;\;\\
\gamma_+^G&=&1/10\;,\;\;
\gamma_+^y=-2/5\;,\;\;
\gamma_-^G=-1/8\;,\;\;
\gamma_-^y=1/2\;,\;\;
\phi=2/9\;,\nonumber
\end{eqnarray}

The singular parts of other thermodynamic variables obey scaling laws
similar to that for $R_{G,y}$, Eq.\ref{Rtrans}. For example, the
singular part of the specific heat per particle 
$C_v\sim{1\over L^D}{\partial^2\over\partial t_y^2}\left({1\over2}t_y
R_y^2 L^{D-2}\right)$, using Eq.\ref{Rtrans} exhibits scaling form
\begin{eqnarray}
C_v&=&L^{\beta} g(t_y L^{\phi})\;,\nonumber\\
&\propto&
\left\{ \begin{array}{lr}
t_y^{-\alpha_+} L^{\beta-\alpha_+\phi}, & t_y > 0,  L >> \xi_{ct} \\
L^{\beta} , & L << \xi_{ct} \\
|t_y|^{-\alpha_-} L^{\beta-\alpha_-\phi}, & t_y < 0, L >> \xi_{ct}
\label{heat}
\end{array} \right.
\end{eqnarray}
with $g(x)\approx {d^2\over d x^2}\left[ f_{y}^2(x)\right]$, and
\begin{eqnarray}
\beta&=&2\nu^y_{ct}-2+\phi\;,\;\;
\alpha_\pm=-2\gamma^y_\pm+1\;,\;\;\\
\beta&=&0\;,\;\;\alpha_+={2d+3\over d+2}\;,\;\;
\alpha_-=0\;,\;\;\;\mbox{Flory theory}.
\label{beta_alpha}
\end{eqnarray}
This leads to the unusual feature that outside the critical regime
(i.e. for $L >> \xi_{ct}$), the singular part of the specific heat
above the crumpled-to-tubule transition vanishes in the thermodynamic
limit like $L^{-\alpha_+\phi}\sim L^{-2(d-D)/(d+2)}\sim L^{-2/5}$.
Similar behavior was also found for the direct crumpled-to-flat 
transition by Paczuski et al..\cite{PKN}

We now turn to the tubule-to-flat (TF) transition. On both sides of
this transition, $R_y=L_y\times O(1)$. Therefore only $R_G$
exhibits critical behavior, which can be summarized by the scaling
law:
\begin{eqnarray}
R_G&=&L^{\nu_{tf}^{\perp,y}} f^{\perp} (t_\perp L^{\phi_{tf}})
\;,\nonumber\\
&\propto&
\left\{ \begin{array}{lr}
t_\perp^{\gamma_{+}^{tf}} L^{\nu_t}, & t_\perp > 0,  L >> \xi_{tf} \\
L^{\nu_{tf}} , & L << \xi_{tf} \\
|t_\perp|^{\gamma_{-}^{tf}} L, & t_\perp < 0, L >> \xi_{tf}
\end{array} \right.
\label{Rtf}
\end{eqnarray}
where $t_\perp=(T-T_{tf})/T_{tf}$, $t_\perp>0$ 
corresponds to the tubule phase,
$\xi_{tf}\propto|t_\perp|^{-1/\phi_{tf}}$ is the correlation length
for this transition, and the exponents obey the scaling relations
%
\begin{eqnarray}
\gamma_{+}^{tf}=(\nu_t-\nu_{tf})/\phi_{tf},\;\;\;
\gamma_{-}^{tf}=(1-\nu_{tf})/\phi_{tf}\;.\label{gamma+-tf1}
\end{eqnarray}
%
In Flory theory we find: 
%
\begin{eqnarray}
\phi_{tf}=1/3,\;\;
\nu_{tf}=5/6,\;\;
\gamma_+^{tf}=-1/4,\;\;
\gamma_-^{tf}=-1/2.\;\;\label{exps}
\end{eqnarray}

The singular part of the specific heat again obeys a scaling law:
\begin{eqnarray}
C_v&=&L^{2\nu_{tf}+\phi_{tf}-2} g(t_\perp L^{\phi_{tf}})\;,\nonumber\\
&\propto&
\left\{ \begin{array}{lr}
t_\perp^{-\alpha_+^{tf}} L^{-\kappa_+}, & t_\perp > 0,  L >> \xi_{tf} \\
L^{2\nu_{tf}+\phi_{tf}-2} , & L << \xi_{tf} \\
|t_\perp|^{-\alpha_-^{tf}} L^{-\kappa_-}, & t_y < 0, L >> \xi_{tf}
\end{array} \right.
\label{cf}
\end{eqnarray}
where, in Flory theory,
%
\begin{eqnarray}
\alpha_+^{tf}=3/2,\;\;
\alpha_-^{tf}=0,\;\;
\kappa_+=1/2,\;\;
2\nu_{tf}+\phi_{tf}-2=\kappa_-=0.\;\;\label{exps2}
\end{eqnarray}

Thus, again, the singular part of the specific heat vanishes (now like
$L^{-1/2}$) in the thermodynamic limit above (i.e., on the tubule
side of) the transition, while it is $O(1)$ and smooth as a function
of temperature in both the critical regime and in the flat phase.

\section{Random Heterogeneity in Polymerized Membranes}
\label{heterogeneity}
\subsection{Motivation}

Soon after a general picture of idealized homogeneous membranes was
established, much of the attention turned to effects of random
heterogeneity on conformational properties of polymerized
membranes.\cite{RN} As with many other condensed 
matter systems (e.g., random magnets, pinned charged density waves,
vortex lattices in superconductors)\cite{otherCMdisorder} 
a main motivation is that random inhomogeneity is an
inevitable feature in most physical membrane realizations. As
illustrated a cartoon of cellular membrane wall (Fig.1.4 in lectures by
Stan Leibler), functional proteins,
nanopores (controlling ionic flow through membrane), and other heterogeneities
(with sincere apologies to biologists for such crude physicist's
terminology) are incorporated into a cellular lipid bilayer.  Holes or tears, 
random variation in the local coordination number (disclinations),
dislocations, grain-boundaries, and impurities incorporated into 
fishnet-like biopolymer spectrin network attached to cellular wall
(as e.g., in red blood cells) are also substantial sources of
inhomogeneity in membranes.  Such defects in the two-dimensional 
polymer network are also inadvertently introduced during 
photo-polymerization of 
synthetic polymerized membranes. If one is only interested in statistical
conformational properties of such membranes, these frozen-in
heterogeneities can be treated as random quenched disorder, similar to
treatment of impurities in other condensed matter contexts.\cite{otherCMdisorder}
%
%

Consistent with theoretical predictions\cite{RN}, 
the fact that quenched internal disorder can drastically modify the
conformational thermodynamics of polymerized membranes was first demonstrated 
experimentally by Mutz, Bensimon and Brienne\cite{Mutz}. 
They observed that partially (heterogeneous) polymerized vesicles
undergo upon cooling a ``wrinkling'' transition to a folded 
rigid glassy stucture that resembles a raisin. Natural 
interpretation of this important experiment as an evidence for a 
transition towards a crumpled spin-glass-like state provided strong
motivation for further theoretical studies of quenched disorder in
polymerized membranes.

Below we will describe a generalized model that
includes effects of quenched disorder in a polymerized membrane
and will show that (as in-plane anisotropy, discussed in previous section)
it has drastic qualitative effects on long-length conformational 
properties of a polymerized membrane. 

\subsection{Model of a heterogeneous polymerized membrane}
\label{model_disorder}

\begin{figure}[bth]
\centering
\setlength{\unitlength}{1mm}
\begin{picture}(150,70)(0,0)
\put(-20,-53){\begin{picture}(150,70)(0,0)
\includegraphics{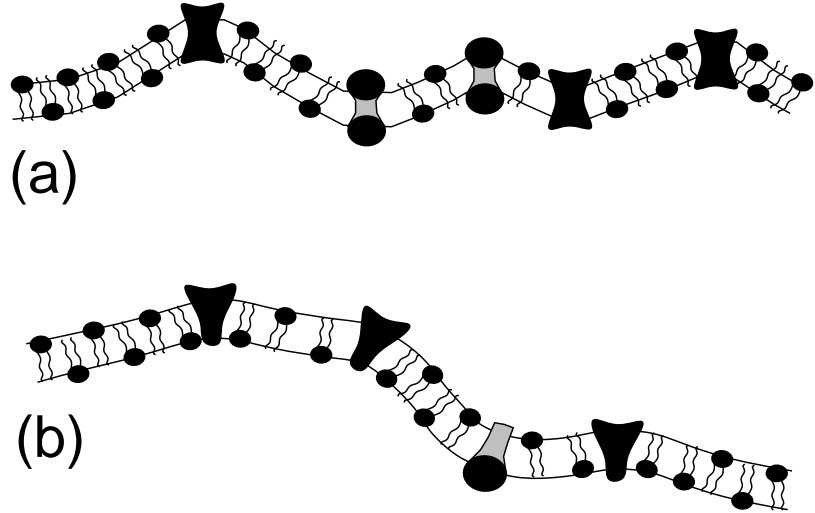}
\end{picture}}
\end{picture}
\caption{A cartoon of a bilayer membrane with a 
reflection- (a) symmetric and (b) asymmetric 
inclusions that can be modelled by two qualitatively distinct types of
disorder, the random stress and random mean curvature disorder, respectively}
\label{symasymDisorder}
\end{figure}

It is clear that above sources of heterogeneity lead to local
random in-plane dilations and compressions and can therefore be
modelled by local random stresses $-\sigma_{\alpha\beta}({\bf x})
u_{\alpha\beta}$.  Geometrically this can be understood as a random
preferred background metric 
$g^0_{\alpha\beta}({\bf x})=
\delta_{\alpha\beta}+\eta^0_{\alpha\beta}({\bf x})$
with strain $\tilde{u}_{\alpha\beta}={1\over2}(g_{\alpha\beta}-
g^0_{\alpha\beta}({\bf x}))$ measured relative to this deformed state, and 
metric $g_{\alpha\beta}=\partial_\alpha{\vec r}\cdot
\partial_\beta{\vec r}$ seeking to relax to $g^0_{\alpha\beta}({\bf
  x}))$. These local in-plane stresses can and will be relaxed by buckling
of the membrane into the third dimension that will tend to screen the
elastic interaction, thereby lowering the elastic energy by partial
trade-off between in-plane elastic energy and membrane bending
(curvature $\kappa$) energy.  However, as illustrated in
Fig.\ref{symasymDisorder}(a), because such randomness
respects the reflection symmetry relating two sides of the
membrane, the induced puckering will locally break this 
Ising symmetry  {\em spontaneously}, and in a way specific to each 
configuration of disorder. 

A qualitatively distinct form of quenched disorder that {\em explicitly} breaks
reflection symmetry arises from asymmetric inclusions of the type
illustrated in Fig.\ref{symasymDisorder}(b). These lead to a local 
preferred extrinsic curvature, that, in the flat phase is described by
$-\nabla^2{\vec h}\cdot{\vec c}({\bf x})$. 

Membrane defects will also of course lead to heterogeneous elastic
moduli, $\kappa({\bf x})$, $\mu({\bf x})$, and $\lambda({\bf x})$. 
However, it can be shown that such weak heterogeneity (i.e., as long
as the average value of these elastic constants remains larger than
their variance, that is they are predominantly positive) has no
qualitative effects on membrane long-scale
conformations. Consequently, all effects of membrane heterogeneity can
be modelled by just two types of quenched disorder, random stress and
random curvature. This is perhaps not surprising given the
aforementioned analogy of membranes with ferromagnet (with membrane
normal $\hat n$ playing the role of a spin $\vec S$), where too, random
bond (that respects ${\vec S}\rightarrow -{\vec S}$) and random field
(that  is odd under ${\vec S}\rightarrow -{\vec S}$) are the only two
qualitatively important types of quenched disorder. One qualitatively important
distinction from random magnets that we can already anticipate at this
point is that the curvature disorder, that couples to the gradient of
the order parameter $\hat n$ is far weaker perturbation than its
ferromagnetic analog, the random field disorder that couples directly
to magnetization. This distinction will lead to importance of the
curvature disorder below $D_{uc}=4$, (same as the random stress disorder and
therefore competing with it) contrasting with the upper-critical
dimension of $D_{uc}=6$ of the random-field in a 
ferromagnet.\cite{otherCMdisorder}

The general model of heterogeneous membrane is therefore described by
an effective Hamiltonian $F[u_\alpha,{\vec h}]$
\begin{equation}
F[u_\alpha,{\vec h}]=
\int d^Dx\left[{\kappa\over2}\big(\nabla^2{\vec h}-
{{\vec c}({\bf x})\over\kappa}\big)^2
+ \mu(u_{\alpha\beta})^2 + {\lambda\over2}(u_{\alpha\alpha})^2 
- \mu u_{\alpha\beta}\eta_{\alpha\beta}({\bf x}) 
- {\lambda\over 2} u_{\beta\beta}\eta_{\alpha\alpha}({\bf x}) \right]\;,
\end{equation}
where quenched disorder fields ${\vec c}({\bf x})$ 
and $\eta_{\alpha\beta}({\bf x})$ can be characterized by zero-mean,
Gaussian statistics with second moment given by:
\begin{eqnarray}
\overline{c_i({\bf x}) c_j({\bf 0})}&=&\delta_{i j}{\Delta}_c({\bf x})\;,\\
\overline{\eta_{\alpha\beta}({\bf x}) \eta_{\gamma\delta}({\bf 0})}&=&
(\Delta_1({\bf x})-{1\over D}\Delta_2({\bf x})) 
\delta_{\alpha\beta}\delta_{\gamma\delta}+
{1\over2}\Delta_2({\bf x})(\delta_{\alpha\gamma}\delta_{\beta\delta}+
\delta_{\alpha\delta}\delta_{\beta\gamma})\;.
\label{variances}
\end{eqnarray}

Another useful form of this model is obtained after the phonons
$u_\alpha$ are integrated out, which, at long length scales, can be
done exactly since the phonons (unlike $\vec h$) are not soft and
therefore can be approximated by a harmonic elasticity. 
The resulting Hamiltonian is given by
\begin{eqnarray}
F[{\vec h}]&=&
\int d^Dx\left[{\kappa\over2}\big(\nabla^2{\vec h}-
{{\vec c}({\bf x})\over\kappa}\big)^2
+{1\over8}\partial_\alpha{\vec h}\cdot\partial_\beta{\vec h}
\left\{2\mu P^T_{\alpha\gamma} P^T_{\beta\delta}
+{2\mu\lambda\over2\mu+\lambda}P^T_{\alpha\beta} P^T_{\gamma\delta}\right\}
\partial_\gamma{\vec h}\cdot\partial_\delta{\vec h}\right.\nonumber\\
&&\left.-{1\over4}\eta_{\alpha\beta}({\bf x})
\left\{2\mu P^T_{\alpha\gamma} P^T_{\beta\delta}
+{2\mu\lambda\over2\mu+\lambda}P^T_{\alpha\beta} P^T_{\gamma\delta}\right\}
\partial_\gamma{\vec h}\cdot\partial_\delta{\vec h}\right]
\end{eqnarray}
where $P^T_{\alpha\beta}=\delta_{\alpha\beta} - 
{\partial_\alpha\partial_\beta\over\nabla^2}$ and 
$P^L_{\alpha\beta}= {\partial_\alpha\partial_\beta\over\nabla^2}$ 
are transverse and longitudinal projection operators.
For $D=2$, $d=3$ membrane $F[{\vec h}]$ simplifies considerably to:
\begin{eqnarray}
F[{\vec h}]=\int d^2x\left[{\kappa\over2}\big(\nabla^2 h-
{c({\bf x})\over\kappa}\big)^2
+{K\over8}\left(P^T_{\alpha\beta}\partial_\alpha h\partial_\beta h
-P^T_{\alpha\beta}\eta_{\alpha\beta}({\bf x})\right)^2\right]\;,
\end{eqnarray}
For a generic configuration of impurity disorder, the ground state is highly
nontrivial as it is determined by simultaneous, but generically 
conflicting, minimization of the extrinsic and Gaussian
curvature ($R={1\over2}((\nabla^2 h)^2 - (\partial_\alpha\partial_\beta h)^2)$)
terms
\begin{eqnarray}
\nabla^2 h = {1\over\kappa} c({\bf x})\;,\ \ \ 
P^T_{\alpha\beta}\partial_\alpha h\partial_\beta h
=P^T_{\alpha\beta}\eta_{\alpha\beta}({\bf x})\;.
\end{eqnarray}
Long-scale properties of such ground state are amenable to statistical
treatment, utilizing standard field theoretic machinery.

\subsection{Weak quenched disorder: ``flat-glass''}
\label{weak_disorder}
\subsubsection{Short-range disorder}
\label{flat_glassSR}

For many (but not all; see below) 
realizations of heterogeneity discussed above, such as for
example random membrane inclusions, the disorder
is short-ranged, and therefore can be characterized by
$\delta$-function correlated disorder with variances
$\Delta_c({\bf x})=\Delta_c \delta^D({\bf x})$, 
$\Delta_{1,2}({\bf x})=\Delta_{1,2} \delta^D({\bf x})$.

To understand the effects of quenched disorder it is helpful to first
study the stability of the flat phase (described by the anomalous
elastic fixed point, studied in Sec.\ref{flat_and_crumpled}) 
by performing a simple 
perturbative calculation in disorder and elastic nonlinearities
directly for a physical membrane ($D=2,d=3$). Standard analysis\cite{RN} then
leads to disorder and thermally renormalized bending rigidity 
$\kappa_R^D$:
\begin{eqnarray}
\kappa^D_R(q)&=&\kappa+
(k_BT\kappa+\Delta_c)\int{d^2p\over(2\pi)^2}
{K[\hat{q}_\alpha P^T_{\alpha\beta}({\bf p})\hat{q}_\beta]^2
\over\kappa^2|{\bf q}+{\bf p}|^4}\nonumber\\
&&-(\Delta_1+\Delta_2)\int{d^2p\over(2\pi)^2}\;
{K^2[\hat{q}_\alpha P^T_{\alpha\beta}({\bf p})\hat{q}_\beta]^2
\over4\kappa|{\bf q}+{\bf p}|^4}\;,
\label{kappaPT}
\end{eqnarray}
The first, temperature-dependent correction that enhances $\kappa$ is
identical to that of a homogeneous membrane and is responsible for the 
stability of the flat phase of polymerized disorder-free membranes.\cite{NP} 
At low temperature the temperature-independent
contributions dominate. The last, random stress contribution leads to a
divergent softening of the bending rigidity,\cite{RN} while the random
curvature term  works to increase the bending rigidity\cite{RN,ML}, 
and thereby
works to stabilize the flat phase through the ``order-from-disorder''
mechanism that is the zero-temperature analog of the thermal one 
discussed in Sec.\ref{flat_and_crumpled}.

Weak disorder should {\it not} affect the asymptotic behavior of membranes in
the flat phase at sufficiently high temperatures, despite its
importance at $T=0$. To see this, assume the disorder is so weak that we can
replace the elastic constants on the right hand side of Eq.\ref{kappaPT} 
by wave-vector-dependent quantities $\kappa_R(p)$ and $K_R(p)$ 
renormalized only by thermal fluctuations in the way controlled by the
disorder-free flat phase fixed point.  As discussed in
Sec.\ref{flat_and_crumpled} these are expected to be singular at long 
scales, with $\kappa_R(p)\sim p^{-\eta_\kappa}$ and $K_R(p)\sim
p^{\eta_u}$.\cite{NP,AL,DG,LRscsa} The expression for $\kappa_R^D(q)$, 
Eq.\ref{kappaPT} becomes
\begin{eqnarray}
\kappa^D_R(q)&=&\kappa_R(q)+
\Delta_c\int{d^2p\over(2\pi)^2}
{K_R({\bf p})[\hat{q}_\alpha P^T_{\alpha\beta}({\bf p})\hat{q}_\beta]^2
\over\kappa_R^2({\bf q}+{\bf p})|{\bf q}+{\bf p}|^4}\nonumber\\
&&-(\Delta_1+\Delta_2)\int{d^2p\over(2\pi)^2}\;
{K_R^2({\bf p})[\hat{q}_\alpha P^T_{\alpha\beta}({\bf p})\hat{q}_\beta]^2
\over4\kappa_R({\bf q}|{\bf q}+{\bf p}|^4}\;,\\
&=&\kappa_R(q)\left[1+{\rm const.}\Delta_c q^{\eta_\kappa}
- {\rm const.}(\Delta_1+\Delta_2) q^{\eta_u}\right]\;,
\label{kappaR}
\end{eqnarray}
where we made use of the exact 2D exponent relation\cite{NP,AL,DG,LRscsa} 
$2\eta_f+\eta_u=2$ (a consequence of rotational invariance). 
Since $\eta_\kappa$ and $\eta_u$ are positive, at finite temperature
weak quenched disorder just gives a {\em subdominant} nonanalytic
correction to disorder-free result for $\kappa_R(q)$.
Physically this finite temperature irrelevance of disorder
comes from singularly soft in-plane elastic moduli and divergent
bending rigidity, that, respectively facilitate
screening of the in-plane stress induced by impurities and suppress
wrinkling effects of curvature disorder.

\begin{figure}[bth]
  \centering \setlength{\unitlength}{1mm}
\begin{picture}(150,60)(0,0)
\put(-20,-103){\begin{picture}(150,0)(0,0)
\includegraphics{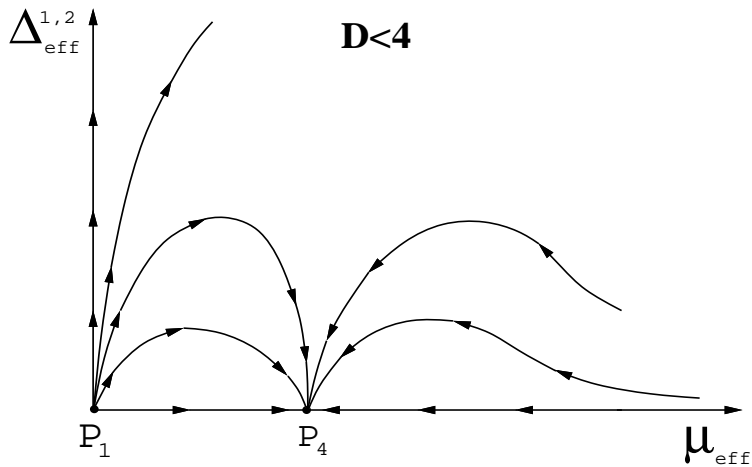}
\end{picture}}
\end{picture}
\caption{RG flow diagram showing the irrelevance of random strain
disorder (disorder variance scales to zero at long scales) 
near the disorder-free finite temperature fixed point P4
controlling properties of the flat phase, and disorder induced
instability of the flat phase at vanishing temperature
($\mu_{eff}\propto\mu T$).}
\label{flowTDelta}
\end{figure}

These perturbative arguments are supported by detailed renormalization
group calculations controlled by an $\epsilon=4-D$-expansion that show
that at finite temperature the Aronovitz-Lubensky fixed point is
stable to weak quenched disorder.\cite{RN} This is summarized by the
RG flow equation of the coupling constants illustrated in Fig.\ref{flowTDelta}

However, as is clear from the flow diagram, for sufficiently 
low-temperatures, even weak bare strain disorder 
becomes strong ($\Delta$'s flow
to large values), invalidating above perturbative argument. In this
case a full low temperature RG analysis is necessary. As first
discovered by Morse, Lubensky and Grest, it shows\cite{LRscsa,RN,ML} that 
interplay of random stress and curvature 
disorder leads to a new stable zero-temperature fixed point that
controls long-scale properties of the disorder-roughened 
polymerized membrane. Similar to the thermally rough flat phase
described by the AL fixed point, the resulting $T=0$ phase is
characterized by a power-law roughness (with $\zeta< 1$) about 
on average flat configuration. It therefore has all the ingredients of
the ``flat-glass'' phase anticipated by Nelson and Radzihovsky.\cite{RN}

These RG results can be nicely complemented by the SCSA analysis.\cite{LRscsa} This
can be done most effectively by first applying the replica
formalism\cite{EA}, that allows one to work with a translationally 
invariant effective Hamiltonian.
To do this, one introduces $n$ copies of fields ${\vec h}^a$ and 
$u_\alpha^a$ labeled by the replica index $a$, and integrates out 
the quenched disorder, thereby obtaining a replicated Hamiltonian. 
Assuming commutability of the thermodynamic and the 
$n\rightarrow 0$ limits,  the relation to the disorder-averaged 
free energy is established through the identity
\begin{equation}
\ln Z=\lim_{n\to 0}{Z^n-1\over n}\;\;,
\end{equation}
where $Z$ the partition function. 

The membrane roughness is characterized by the full disorder-averaged 
height correlation function (that can also be related to replicated ones):
\begin{eqnarray}
\overline{\langle({\vec h}({\bf x})-{\vec h}({\bf 0}))^2\rangle}&=&
\overline{\langle({\vec h}({\bf x})-{\vec h}({\bf 0}))^2\rangle}_{\rm conn}+
\overline{\langle{\vec h}({\bf x})-{\vec h}({\bf 0})\rangle^2}\;,\\
&\sim& A_c|{\bf x}|^{2\zeta} + A|{\bf x}|^{2\zeta'} \;,
\end{eqnarray}
where, respectively, the first (connected) and second contributions 
characterize thermal- and disorder-generated roughness, with
corresponding roughness exponents $\zeta$ and $\zeta'$, and 
the overbar denotes configurational (disorder) averages. The related
exponents characterizing the Fourier transform of these parts of the
height correlation functions are given by $\zeta=(4-D-\eta_\kappa)/2$ and 
$\zeta'=(4-D-\eta'_\kappa)/2$. Analysis very similar to that done for
homogeneous membranes in Sec.\ref{SCSAflat}, but with additional replica
matrix structure leads to a zero-temperature fixed point, that is 
marginally unstable to finite temperature.\cite{ML} It is characterized by
$\eta_\kappa=\eta'_\kappa$, with 
\begin{equation}
\eta_\kappa(d_c,D) = \eta_\kappa^{\rm pure}(4 d_c,D)\;,
\end{equation}
where $\eta^{\rm pure}_\kappa(d_c,D)$ is the SCSA exponent for
$\eta_\kappa$ found
in Sec.\ref{SCSAflat} characterizing a homogeneous polymerized membrane at
a finite temperature. The underlying
reasons for this amazing connection between roughness exponents at the
disorder- and thermally-dominated fixed points is unclear. However, this SCSA
prediction agrees with the $1/d_c$- and $\epsilon$-expansions to lowest
order in respective small parameters.  For a physical membrane, $D=2$, 
$d_c=1$, SCSA predicts:
\begin{equation}
\eta= 2/(2+\sqrt{6})=0.449\;,\;\;\;\zeta=0.775\;,
\end{equation}
that compares quite well (and much better than the lowest order 
$\epsilon$- or $1/d_c$-expansions) with the numerical simulation\cite{ML} 
result $\zeta=0.81\pm 0.03$ for a heterogeneous polymerized membrane.

\subsubsection{Long-range disorder}
\label{flat_glassLR}

\begin{figure}[bth]
  \centering \setlength{\unitlength}{1mm}
\begin{picture}(150,80)(0,0)
\put(-10,-70){\begin{picture}(150,0)(0,0)
\includegraphics{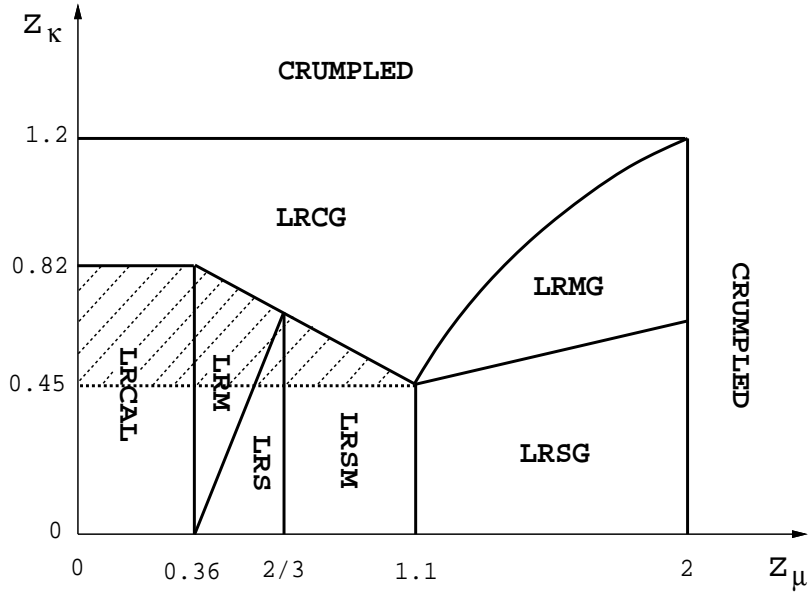}
\end{picture}}
\end{picture}
\caption{Domain of stability of the flat phases as a function of
  $z_\mu,z_\kappa$. 
(1) Disorder-dominated phases ($\zeta'>\zeta$):
long-range stress glass (LRSG), long-range curvature glass (LRCG), 
long-range mixed glass (LRMG).
(2) Temperature-dominated phases ($\zeta'<\zeta$): LR curvature
(LRCAL), 
LR mixed (LRM) and LR stress (LRS). (3) LRSM: marginal phase with
$\zeta=\zeta'$.
The shaded area corresponds to a region of thermal phase transitions 
between several stable phases (LRCG and the others). The region where 
the membrane crumples is indicated.}
\label{disorderLR}
\end{figure}

Above analysis of short-range impurity disorder can be easily extended
to treat disorder with long-range correlations, that can arise from weakly
correlated distribution of frozen-in dislocations and disclinations,
random grain boundaries\cite{NRgrainboundary},
and from polymerized-in quasi-long-range correlated lipid tilt (or other
membrane vector) order. At long scales, such disorder can be
characterized by variances with power-law Fourier transforms:
\begin{eqnarray}
\Delta_c({\bf q})&=&\Delta_c q^{-z_c}\;,\\
\Delta_{1,2}({\bf q})&=&\Delta_{1,2} q^{-z_{1,2}}\;,
\end{eqnarray}
where $z_c$ and $z_{1,2}$ are curvature and stress disorder 
correlation exponents. Such long-range disorder considerably enriches
the phase diagram of heterogeneous polymerized membranes, introducing
a number of new flat-glass phases, that are summarized
as function of value of these range exponents in Fig.\ref{disorderLR}.
For sufficiently short-ranged disorder (both $z_c$ and $z_{1,2}$
small), and for finite and zero temperature, we respectively recover
the SCSA exponents for the Aronovitz-Lubensky\cite{AL,DG,LRscsa} and 
Morse-Lubensky\cite{ML,LRscsa} fixed points. 

More generally, the nature of the stable phases strongly depends on the value of the
disorder-range $z$ exponents and divides into three classes: (1) $\zeta >
\zeta'$ with temperature dominated roughness, (2) $\zeta <
\zeta'$ with disorder dominated roughness, and (3) $\zeta =\zeta'$ 
with equal scaling of the disorder and thermal contributions to the
membrane roughness. Each one of these regions then further subdivides
into distinct flat-glass phases depending on whether long-range curvature,
stress, or both types of disorders are relevant. 
That is, in the presence of long-range disorder four new flat-glass 
phases, stable at finite temperature appear:
(i) short-range (SR) curvature and long-range (LR) stress disorder
(LRSG in Fig.\ref{disorderLR}), (ii) LR curvature and SR stress disorder (LRCG),
(iii) LR disorder in both curvature and stress disorder (LRMG), 
and (iv) zero curvature disorder and LR stress disorder (not
represented in Fig.\ref{disorderLR}). In addition to these flat-glass phases 
three corresponding temperature-dominated flat phases (LRS, LRCAL, LRM), and
two phases for which $\zeta=\zeta'$ appear. In the shaded area in 
Fig.\ref{disorderLR} several 
of these phases are stable. Phase transition controlled by
strength of disorder and/or temperature is therefore  expected between
them.

For sufficiently long range disorder correlations (large $z$'s), the
dominant roughness exponent $\zeta'$ reaches $1$, presumably
indicating disorder-driven crumpling transition, and therefore
breakdown of the weak-disorder expansion about (on average) flat phase.
The expected energy-driven crumpling instability to a qualitatively
distinct {\em isotropic} ``crumpled-glass'' state is a priori of entirely
different nature than the entropy-driven crumpling
transition predicted for phantom membranes.

\subsection{Strong quenched disorder: ``crumpled-glass''}
\label{crumpled_glassSec}

Description of the strongly disordered crumpled-glass phase and the
associated transition is significantly more complicated, because in
addition to complexities of conventional spin-glasses, nonlocal
self-avoiding interaction must be included. Such crumpled
phase is characterized by a vanishing average tangent field 
$\overline{\langle\partial_\alpha
  r_i\rangle}$ (hence crumpled), but with a nonzero crumpled-glass
order parameter $\overline{\langle\partial_\alpha r_{i}\rangle \langle \partial_\beta
  r_{j}\rangle}$ analogous to Edwards-Anderson spin-glass order
parameter in disordered magnets.\cite{otherCMdisorder}
Some progress toward description of such crumpled-glass 
phase in phantom membranes 
was made by Radzihovsky and Le Doussal\cite{RLlarge_d}, 
by utilizing a $1/d$-expansion.\cite{largeN,largeNcomment,ZinnJustin}

In the limit of large embedding dimension, $d\rightarrow\infty$, the
{\it homogeneous} membrane model can be solved exactly.\cite{DG}
In contrast, for a {\it disordered} membrane even in the 
$d\rightarrow\infty$ limit the exact solution of the crumpled-glass 
phase appears to be intractable. The difficulty arises from the tensor
structure of the crumpled-glass 
order parameter, that leads to a problem of matrix
field theory, a notoriously difficult problem.  However, some progress
can be made within an additional mean-field like approximation that
ignores fluctuations in the tensor crumpled-glass order parameter. 
To simplify technical aspects of the presentation it is 
convenient to specialize to
a purely scalar stress-only disorder, described by random Gaussian, zero-mean
dilations and compressions in the locally preferred metric, 
$g^0_{\alpha\beta}({\bf x})=\delta_{\alpha\beta}(1+\delta g^0({\bf
  x}))$. A much more questionable (but technically necessary)
approximation is omission of the self-avoiding interaction, that is
undoubtedly important in the crumpled-glass phase. 

Because of the isotropic nature of the crumpled-glass phase, the model must be
formulated in terms of $d$-dimensional conformation vector ${\vec
  r}({\bf x})$.  The effective Hamiltonian is given by
\begin{eqnarray}
\hspace{-0cm}F[{\vec r}]=d \int d^Dx\left[{\kappa\over2}|\nabla^2{\vec r}|^2
+{\mu\over4}\big(\partial_\alpha{\vec r}\cdot\partial_\beta{\vec r}
-g^0_{\alpha\beta}({\bf x})\big)^2
+{\lambda\over8}\big(\partial_\alpha{\vec r}\cdot\partial_\alpha{\vec r}-
g^0_{\alpha\alpha}({\bf x})\big)^2\right]\hspace{0.3cm}\;\;\;\;\hfill
\label{Fcrumpled}
\end{eqnarray}
where the elastic moduli were rescaled by $d$  so as to obtain
sensible and nontrivial results in the limit $d\rightarrow\infty$.  
Replicating $F$ allows averaging over quenched
disorder.\cite{EA} Then, introducing two Hubbard-Stratanovich fields
$\chi_{\alpha\beta}$ and $Q_{a b\alpha\beta}^{ij}$ to 
decouple replica diagonal and off-diagonal nonlinearities, respectively, leads to
an effective Hamiltonian that is quadratic in $\vec r$. This allows
formal integration over $\vec r$, that is conveniently done around background
configuration ${\vec r}_0$. Now, ignoring fluctuations in the
Hubbard-Stratanovich fields, the values of the order parameters
$\partial_\alpha{\vec r}_0$ and $Q_{a b\alpha\beta i j}^{o}$ are
determined by minimizing the resulting replicated free
energy, together with the equation of
constraint relating $\chi_{a\alpha\beta}$ to these order parameters.

Assuming that the replica symmetry breaking does not occur until
higher order in $1/d$, as it happens in the random anisotropy axis
model\cite{Goldschmidt,KJK} we look for the saddle point
replica-symmetric solution of the following form,
\begin{eqnarray}
\vec{r}^o_{a}=\zeta x_\alpha \hat e_\alpha\;,\;\;\;
\chi^o_{a\alpha\beta}=\chi\delta_{\alpha\beta}\;,\;\;\;
Q_{ab\alpha\beta ij}^o=Q\delta_{\alpha\beta}\delta_{ij}(1-\delta_{ab})\;.
\label{ansatz_d}
\end{eqnarray}
The corresponding saddle-point equations for $\zeta$, $\chi$ and $Q$ are given by:
\begin{eqnarray}
(1-\zeta^2)+2\chi(\alpha+\beta D)&=&{T\over 2D}\int_0^\Lambda{d^Dk\over(2\pi)^D}
\left[{2\over\kappa k^2+\chi+{\hat\Delta\over 2T} Q}+{\hat\Delta Q/T
\over(\kappa k^2+\chi+{\hat\Delta\over 2T}Q)^2}\right],\hspace{0.7cm}
\label{saddleEq1}\\
\zeta^2&=&Q\left(1-{{\hat\Delta}\over2D}\int_0^\Lambda{{d^Dk}\over{(2\pi)^D}}
{1\over(\kappa k^2+\chi+{\hat\Delta\over 2T}Q)^2}\right)\;,
\label{saddleEq2}\\
\left(\chi+{Q\hat\Delta\over2T}\right)\zeta&=&0\;,
\label{saddleEq3}
\end{eqnarray}
where $\Delta$ is scalar stress disorder variance and
$\hat\Delta=(2\mu + D\lambda)\Delta$. For the special
case of homogeneous membranes, $\Delta=0$, these equations
reassuringly reduce to those found by David, et al.\cite{DG}
and describe the crumpled-to-flat transition exactly to leading order in
$1/d$. 

For a heterogeneous membrane ($\hat\Delta > 0$) there are three distinct
solutions to these saddle-point equations, corresponding to three different
possibilities for the values of the pair of order parameters $\zeta$
and $Q$. 
\begin{eqnarray}
\zeta&=&0\;,\;\;Q=0\;\;,\\
\zeta&\neq& 0\;,\;\;Q\neq 0\;\;,\\
\zeta&=&0\;,\;\;Q\neq 0\;\;.
\end{eqnarray}
that correspond to the crumpled phase, flat phase and crumpled-glass
phase of the membrane, respectively. 

\begin{figure}[bth]
  \centering \setlength{\unitlength}{1mm}
\begin{picture}(150,60)(0,0)
\put(-5,-85){\begin{picture}(150,0)(0,0)
\includegraphics{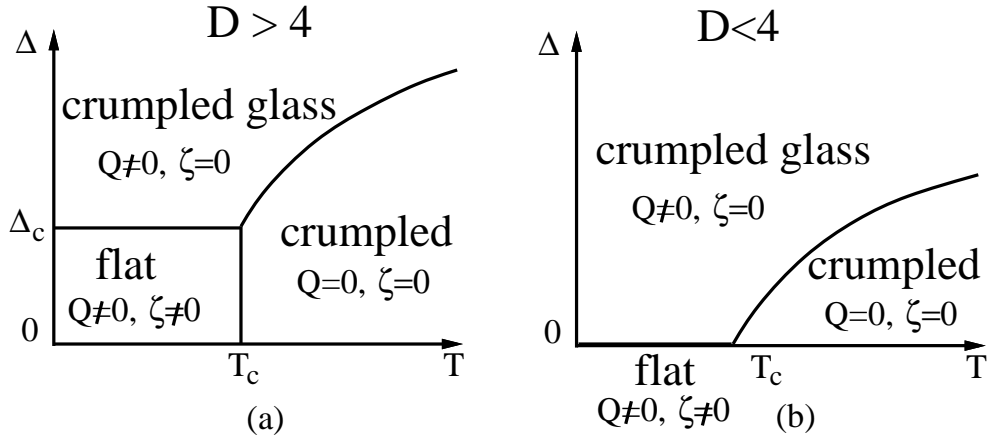}
\end{picture}}
\end{picture}
\caption{A phase diagram for a disordered polymerized membrane 
showing (a) crumpled, flat- and crumpled-glass phases for $D>4$. (b) In
the physical case $D=2<4$ the flat phase does not appear 
in this $d\rightarrow\infty$ limit, but is
expected to when $1/d$ corrections are taken into account.}
\label{crumpled_glass}
\end{figure}

Critical properties of these three phases and 
phase boundaries between them can be obtained from a straightforward
analysis of Eqs.\ref{saddleEq1}-\ref{saddleEq3}.\cite{RLlarge_d} 
The phase behavior is summarized in 
Fig.\ref{crumpled_glass}, illustrating that within this approximation
the lower-critical dimension for the flat phase in the presence of
quenched disorder is $D_{lc}^\Delta=4$, and
therefore for this model (in $d\rightarrow\infty$ limit) only crumpled-glass
and thermal crumpled phases survive in $D=2$ membranes. 

For the flat phase saddle point equations give 
\begin{eqnarray}
\zeta^2=A\left(1-{T\over T_c}\right)\left(1-
{\Delta\over\Delta_c}\right)\;\;,\;\;\;
Q={A}\left(1-{T\over T_c}\right)\;\;.
\end{eqnarray}
where critical crumpling transition temperature $T_c$ and critical value of
disorder $\Delta_c$ (not to be confused with the variance of the 
curvature disorder from
Secs.\ref{model_disorder}-\ref{flat_glassLR}), 
defining the rectangular boundaries of the flat phase, are given by 
\begin{eqnarray}
T_c^{-1}={1\over D}\int_0^\Lambda{d^Dk\over(2\pi)^D}
{1\over\kappa k^2}\;,\;\;\;
\Delta_c^{-1}={1\over2D}\int_0^\Lambda{{d^Dk}\over
{(2\pi)^D}}{1\over{\kappa^2k^4}}\;,
\end{eqnarray}
and $A^{-1}=1+(\alpha+D\beta){\hat\Delta}/T$. The power-law vanishing of
these order parameters according $\zeta\sim(T_c-T)^{\beta^\zeta_T}$,
$\zeta\sim(\hat\Delta_c-\hat\Delta)^{\beta^\zeta_\Delta}$ and
$Q\sim(T_c-T)^{\beta^Q_T}$, defines the corresponding $\beta$
exponents: $\beta^\zeta_T={1/2}$, $\beta^\zeta_\Delta=1/2$, 
$\beta^Q_T=1$. 

Outside this rectangular region $\zeta$ vanishes and the membrane undergoes a
crumpling transition out of the flat phase. For 
$\hat\Delta\leq\hat\Delta_c$ and as 
$T\rightarrow T_c$ the transition is 
to the crumpled phase, while for $T\leq T_c$ and as
$\hat\Delta\rightarrow\hat\Delta_c$ flat phase is unstable to 
the crumpled-glass phase. 
%


The glass susceptibility near the transition from
the flat phase to the crumpled-glass phase can also be easily
calculated by introducing an external field 
$h_{ij\alpha\beta}=h\delta_{ij}\delta_{\alpha\beta}$ 
conjugate to the crumpled-glass
order parameter $Q_{ab\alpha\beta ij}$. The resulting saddle-point
equations then lead to the crumpled-glass susceptibility,
$\chi_{sg}=\partial Q/\partial h\sim(\hat\Delta_c-\hat\Delta)^{-\gamma_{sg2}}$,
with $\gamma_{sg2}=1$.

Similarly, upon approach to the flat phase from the crumpled-glass
(characterized by $Q\neq 0$ and $\zeta=0$) 
the tangent susceptibility $\chi_\zeta$ diverges as 
tangent order $\zeta$ spontaneously develops. 
Turning on an external field $f$ that couples to the tangent order parameter,
leads to 
$\chi_{\zeta}=\partial\zeta/\partial f\sim(\Delta-\Delta_c)^{-\gamma_{\zeta 2}}$
giving $\gamma_{\zeta 2}=2/|4-D|$, that, as expected diverges at the lower-critical
dimension $D_{lc}^\Delta=4$ of the flat phase.

Finally, we look at the transition between the crumpled and 
crumpled-glass phases.
The crumpled-glass susceptibility $\chi_{sg}$ near this transition is
given by $\chi_{sg}\sim(\Delta_c(T)-\Delta)^{-\gamma_{sg1}}$, with
$\gamma_{sg1}=1$, as at the flat-to-crumpled-glass transition, except
for the modified phase boundary that is nonzero for any $D$:
\begin{equation}
\Delta_c^{-1}(T)={1\over2D}\int_0^\Lambda{{d^Dk}\over
{(2\pi)^D}}{1\over{\left(\kappa k^2+\chi(T)\right)^2}}\;\;,
\end{equation}
and together with saddle-point equations and $Q=0$, also defines the phase 
boundary between the crumpled and crumpled-glass phases for $D>2$,
$\Delta_c(T)-\Delta_c\sim (T-T_c)^\phi$, with 
$\phi$ the crossover exponent $\phi=|D-4|/(D-2)$.

\begin{figure}[bth]
  \centering \setlength{\unitlength}{1mm}
\begin{picture}(150,55)(0,0)
\put(-20,-68){\begin{picture}(150,0)(0,0)
\includegraphics{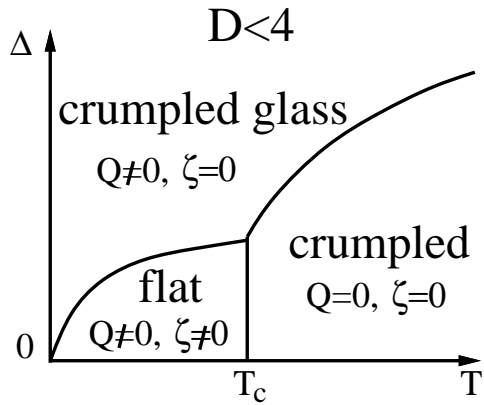}
\end{picture}}
\end{picture}
\caption{Conjectured phase diagram for a disordered membrane $D\leq 4$
  when $1/d$ corrections are taken into account, that allow a region
  of flat phase of size $1/d$ to appear.}  
\label{crumpled_glass_d}
\end{figure}

As noted above, $d\rightarrow\infty$ analysis predicts an instability of the
flat phase of $D=2$ membranes to any amount of disorder
($D_{lc}^\Delta=4$).  A computation of $1/d$ corrections for the disordered
membrane is technically quite challenging and remains an open
problem. However, quite generally, anomalous elasticity generated by $1/d$
corrections (e.g., finite $\eta_\kappa=O(1/d)$ exponent\cite{DG}) 
strongly suggests
the lowering of $D_{lc}^\Delta$  down to $D_{lc}^\Delta=4-O(1/d)$. This is
supported by the Harris criterion applied to the buckling transition,
controlled by the Aronovitz-Lubensky fixed point, that leads to
stability of the flat phase (and the AL fixed point) as long as 
$\eta_u$ remains positive. This is also consistent with the
$\epsilon=4-D$-expansion analysis of Radzihovsky and Nelson (performed
for arbitrary $d$)\cite{RN}, discussed in Sec.\ref{flat_glassSR},
that the lower-critical dimension is reduced down to $D_{lc}^\Delta=4-4/d$.
A phase diagram for $D\leq4$ consistent with the nature of 
the $1/d$ corrections is illustrated in Fig.\ref{crumpled_glass_d}.

Finally we observe that the crumpled-glass phase can be destroyed
by applying an external tension to the membrane's boundaries.  The
metastable degenerate ground states 
would disappear and the average of the local
tangents would no longer vanish. In this respect an external stress would be
analogous to an external magnetic field in spin systems.  
As the stress is reduced the membrane would slowly return to the 
glassy phase but with some hysteresis. The line separating
the regions of stable and metastable degenerate states is then the 
analogue of the d'Almeida-Thouless line\cite{AT} 
studied in great detail for the real spin-glasses.\cite{otherCMdisorder}

\section{Interplay of Anisotropy and Heterogeneity: 
Nematic Elastomer Membranes}

\begin{figure}[bth]
\centering
\setlength{\unitlength}{1mm}
\begin{picture}(150,70)(0,0)
\put(0,-70){\begin{picture}(150,70)(0,0)
\includegraphics{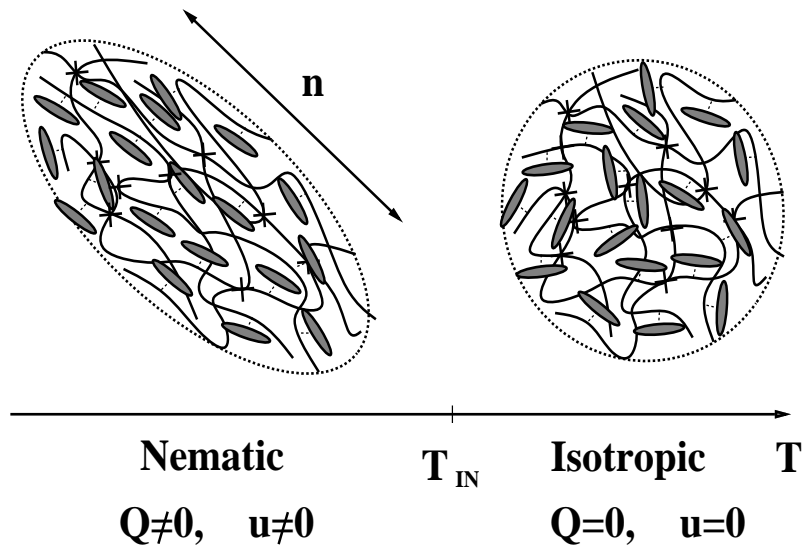}
\end{picture}}
\end{picture}
\caption{Spontaneous uniaxial distortion of nematic elastomer driven by
  isotropic-nematic transition.}
\label{NEtransition}
\end{figure}

We would like to conclude these lectures with a discussion of 
a new exciting realization of polymerized membranes, nematic
elastomer membranes.\cite{XingNEmembrane}  The motivation for their study is 
driven by recent experimental progress in the
synthesis of nematic liquid-crystal elastomers\cite{reviewNE},
statistically isotropic and homogeneous gels of crosslinked polymers
(rubber), with main- or side-chain mesogens, that can {\em
  spontaneously} develop nematic orientational order, accompanied by a
spontaneous uniaxial distortion illustrated in Fig.\ref{NEtransition}.

\begin{figure}[bth]
\centering
\setlength{\unitlength}{1mm}
\begin{picture}(150,60)(0,0)
\put(0,-70){\begin{picture}(150,70)(0,0)
\includegraphics{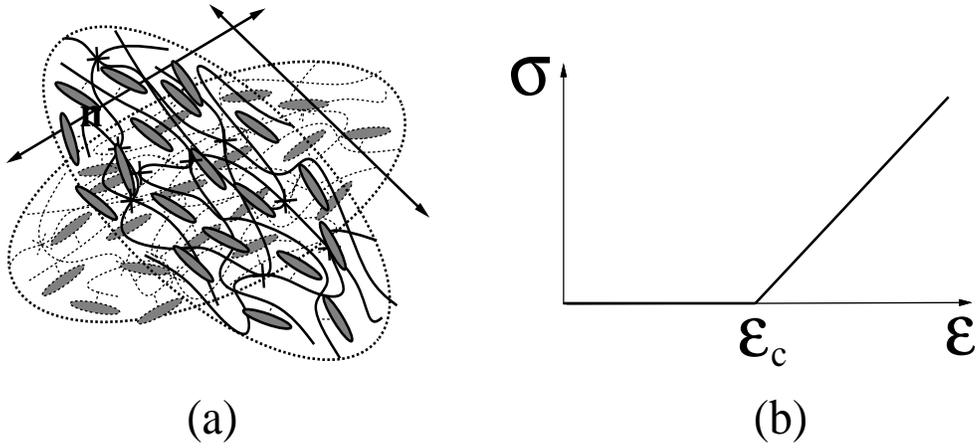}
\end{picture}}
\end{picture}
\caption{(a) Simultaneous reorientation of the nematic director and of the
  uniaxial distortion is a low-energy nemato-elastic Goldstone mode of an ideal
  elastomer, that is responsible for its softness and (b) its flat
  (vanishing stress) stress-strain curve for a range of strains.}
\label{NEgm}
\end{figure}

Even in the absence of fluctuations, {\em bulk} nematic elastomers
were predicted\cite{GL} and later observed to display an array of fascinating
phenomena\cite{reviewNE,LMRX}, the most striking of which is the
vanishing of stress for a range of strain, applied transversely to the
spontaneous nematic direction. This striking softness is generic, stemming from
the {\em spontaneous} orientational symmetry breaking by the nematic
state\cite{GL,LMRX}, accompanied by a Goldstone mode\cite{commentSoft}, 
that leads to the observed
soft distortion and strain-induced director 
reorientation\cite{stress_strainExps}, illustrated
in Fig.\ref{NEgm}. The hidden
rotational symmetry also guarantees a vanishing of one of the five
elastic constants\cite{LMRX}, that usually characterize harmonic
deformations of a three-dimensional uniaxial
solid.\cite{LandauLifshitz} Given the discussion in Sec.\ref{flatAnomElast}, 
not surprisingly, the resulting elastic
softness leads to qualitative importance of thermal fluctuations and
local heterogeneity. Similar to their effects in smectic and columnar liquid
crystals,\cite{GP,RTsmectic,RTcolumnar} thermal fluctuations lead to
anomalous elasticity (universally length-scale dependent elastic
moduli) in bulk homogeneous elastomers with dimensions below
$3$,\cite{NEelXR,NEelSL} and below $5$, when effects of the
random network heterogeneity are taken into account.\cite{NEprlXR}

This rich behavior of bulk elastomers provided strong motivation to
study nematic elastomer membranes ($D$-dimensional sheets of nematic elastomer
fluctuating in $d$ dimensions).\cite{XingNEmembrane}  
A model of such a membrane must
incorporate both network anisotropy and heterogeneity discussed in
previous sections.  However, an important distinction from {\em explicitly}
anisotropic membranes discussed in Sec.\ref{anisotropy} is that the
nematic anisotropy is {\em spontaneously} chosen in the amorphous 
(initially statistically isotropic) elastomer matrix.  At
harmonic level this in-plane rotational symmetry can be captured by a
two-dimensional harmonic effective Hamiltonian
\begin{eqnarray}
{\cal H}^0_{NE} &=& {1\over 2}\int d^2x\big[ \kappa_{xx}(\partial_x^2h)^2
  + \kappa_{yy}(\partial_y^2 h)^2
  + 2\kappa_{xy} (\partial_x^2h)(\partial_y^2h)
  + K_y (\partial_y^2 u_x)^2
   +K_x(\partial_x^2 u_y)^2  \nonumber \\
& & +\lambda_{x}(\varepsilon_{xx})^2 + 
\lambda_y (\varepsilon_{yy})^2 + 
2\lambda_{xy} \varepsilon_{xx}\varepsilon_{yy}\big] ,
\label{Hne0}
\end{eqnarray}
with $\varepsilon_{\alpha\alpha}=\partial_\alpha u_\alpha$ (no sum over
$\alpha$ implied here), and 
characterized by a uniaxial phonon elasticity with a vanishing transverse shear
modulus, $\mu_{xy}$. This latter feature is what distinguishes a 
nematic elastomer membrane from an 
{\em explicitly} anisotropic membrane discussed in Sec.\ref{anisotropy}.  
The vanishing in-plane shear modulus captures at the harmonic
level the invariance of the free energy under infinitesimal rotation
of the nematic axis and the accompanying uniaxial 
distortion of the elastomer matrix.
To ensure an in-plane stability curvature phonon elastic energies are 
included in ${\cal H}^0_{NE}$. 

As a result, in the putative flat nematic phase
of an elastomer membrane, the phonons have qualitatively ``softer''
harmonic elasticity than in a conventional polymerized membrane
discussed in Sec.\ref{anisotropy}.  Consequently, as in other ``soft''
systems (e.g., smectic and columnar liquid crystals phases; see
discussion in Sec.\ref{flatAnomElast}), in the presence of thermal
fluctuations and/or heterogeneities, {\em nonlinear} elastic terms are
essential for the correct description. 

First principles derivation of the nematic 
elastomer model, that incorporates (hidden)
in-plane rotational invariance at nonlinear level is somewhat involved
and we refer an interested reader to recent detailed work on this
subject.\cite{NEelXR,NEelSL}  
In short, one starts out with a model of a statistically
homogeneous and isotropic elastic membrane coupled to a nematic
in-plane order parameter $Q_{\alpha\beta}$. The rotational symmetry is then
spontaneously broken by the nematic ordering at the isotropic-nematic 
transition, that also induces
a spontaneous uniaxial in-plane distortion of the elastomer matrix,
characterized by a strain tensor $u_{\alpha\beta}^0$.  Expansion about
this flat uniaxial state, ensuring underlying in-plane and embedding
space rotational invariance leads to a nonlinear elastic Hamiltonian
of a nematic elastomer membrane. Its form is that of the ${\cal H}_{NE}^0$,
Eq.\ref{Hne0}, but with the harmonic strain $\varepsilon_{\alpha\beta}$
replaced by a nonlinear strain tensor $w_{\alpha\beta}$, 
that incorporates both in-plane
and height nonlinearities of the form $(\partial_x u_y)^2$ and 
$(\partial_x h)^2$, respectively.  

In $D=2$ both phonon and height anharmonic terms are strongly
relevant (their perturbative corrections grow with length scale) and
therefore must be both taken into account. One approach is to
generalize the above model to a $D$-dimensional nematic elastomer membrane
and perform an RG calculation controlled by an $\epsilon$-expansion. 
However, it is easy to show that upper-critical dimensions for these
nonlinearities are different, with height undulations becoming relevant
below $D^h_{uc}=4$, and smectic-like and columnar-like in-plane nonlinearities
with upper-critical dimensions of $D^{sm}_{uc}=3$ and $D^{col}_{uc}=5/2$,
respectively.  Hence for $D>3$, in-plane phonon nonlinearities can
be neglected, with height undulation nonlinearities (of the type
studied in Secs.\ref{SCSAflat},\ref{tubuleAnomalous} the only remaining
relevant ones for $D<4$, and controllable close to $D=4$ with an
$\epsilon=4-D$-expansion.  

Generically, one would expect these undulation nonlinearities to
renormalize bending rigidities $\kappa_{\alpha\beta}$ 
as well as in-plane elastic moduli
$\lambda_{\alpha\beta}$, leading to anomalous elasticity with
$\kappa_{\alpha\beta}({\bf q}) \sim q^{-\eta_\kappa}$, 
$\lambda_{\alpha\beta}({\bf q}) \sim q^{\eta_\lambda}$. As discussed
in Sec.\ref{SCSAflat}, here too, rotational invariance
imposes an exact Ward identity between exponents:
\begin{equation}
2 \eta_{\kappa} + \eta_{\lambda} = \epsilon = 4 - D .
\label{WardNEmembrane}
\end{equation} 
However, it is not difficult to show\cite{XingNEmembrane}, that once
in-plane nonlinearities are neglected (legitimate for $D>3$), the
harmonic phonons can be integrated
out exactly, and lead to a purely harmonic effective Hamiltonian ${\cal
  H}[h]$. Therefore there is a strict {\em non}renormalization of
the bending rigidity tensor $\kappa_{\alpha\beta}$ for $D>3$. This
together with the Ward identity, Eq.\ref{WardNEmembrane} gives 
\begin{equation}
\eta_\kappa = 0,\;\;\; \eta_\lambda=4-D\;,
\end{equation}
a result that is supported by a detailed renormalization group
calculation.\cite{XingNEmembrane} This analysis also makes 
contact and recovers some of the results previously obtained in the studies of
isotropic polymerized membranes. In particular, the previously
seemingly unphysical, the so-called ``connected fluid''\cite{AL} is realized as
a fixed point of a nematically-ordered elastomer membrane that is unstable to the
globally stable nematic-elastomer fixed point.\cite{XingNEmembrane}

Despite of some success, there are obvious limitations of above 
description, most notably in its application to the physical
case of $D=2$ elastomer membranes and inclusion of the (usually more
dominant) local network heterogeneity.  The first shortcoming primarily has to
do with the neglect of in-plane elastic nonlinearities, which, near
the Gaussian fixed point become relevant for $D<3$. While it is very
likely that the subdominance of these in-plane nonlinearities relative
to the undulation ones will persist some amount {\em below} $D=3$
\cite{phi4comment}, we expect that in the physical case of $D=2$ all
three nonlinearities need to be treated on equal footing. Carrying
this out in a consistent treatment remains an open and challenging problem.

Secondly, elastomers are only statistically homogeneous and isotropic,
exhibiting significant local heterogeneity in the polymer
network. As we saw in Sec.\ref{heterogeneity}, such 
internal quenched disorder has rich qualitative
effects in ordinary polymerized membranes. Furthermore, recent work by
Xing and Radzihovsky has demonstrated,\cite{NEprlXR} that interplay between
nonlinear elasticity and random strains and torques 
(due to network heterogeneity) leads to disorder
controlled anomalous elasticity even in three-dimensional bulk nematic
elastomers. Because nematic elastomer membranes
are far softer than ordinary polymerized membranes and their bulk
analogs, we expect network heterogeneity to have
strong and rich effects in these systems. Considerable research remains to 
work out the resulting phenomenology.
\begin{figure}[bth]
\centering
\setlength{\unitlength}{1mm}
\begin{picture}(150,30)(0,0)
\put(-3,0){\begin{picture}(150,30)(0,0)
\includegraphics{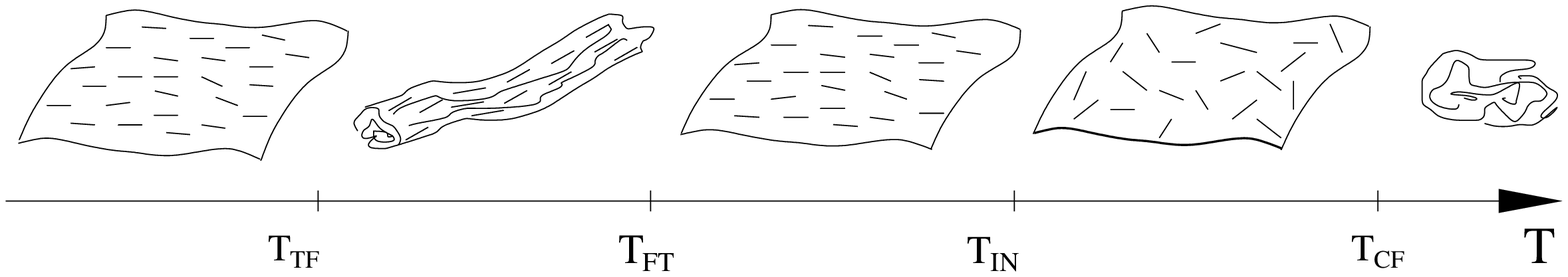}
\end{picture}}
\end{picture}
\caption{A possible phase diagram for ideal nematic elastomer
membranes.  As the temperature is lowered a crumpled membrane
undergoes a transition to isotropic flat phase at $T_{CF}$, followed
by a 2D in-plane isotropic-nematic like transition to an
anisotropic (nematic) flat phase.  As $T$ is lowered further, this
anisotropic flat phase becomes unstable to a nematic tubule
phase, where it continuously crumples
in one direction but remains extended in the other. At even lower
temperature,
a tubule-flat transition takes place at $T_{TF}$.}
\label{NEphasediagram}
\end{figure}

It is interesting to conclude with a general discussion of the global 
conformational phase behavior of nematic elastomer 
membranes.  As with ordinary polymerized
membranes, upon cooling, isotropic elastomer membranes should 
undergo a crumpling (flattening) transition from the crumpled to
flat-isotropic phase. Upon further cooling, an in-plane (flat)
isotropic-to-(flat) nematic transition can take place, leading to a
flat membrane with a spontaneous in-plane nematic order.  As for explicitly
anisotropic membranes discussed in Sec.\ref{anisotropy}, 
such nematically-ordered elastomer membranes should undergo further
transition to a nematically-ordered tubule phase. Because of the
in-plane rotation symmetry that is {\em spontaneously} (as 
opposed to explicitly)
broken, such nematic tubule will exhibit in-plane elasticity (``soft''
phonons) that is qualitatively distinct from tubules discussed in
Sec.\ref{tubuleAnomalous}, and will constitute a distinct phase 
of elastic membranes.\cite{tubule_unpublished} 
One interesting scenario of phase progression is
the nematic-flat to nematic-tubule to nematic-flat reentrant phase
transitions, driven by competition between growth of nematic order
(anisotropy) and suppression of membrane's out-of-plane undulations
upon cooling, as schematically illustrated 
in Fig.\ref{NEphasediagram}. 
Considerable research to elucidate the nature of the 
resulting fascinating phases and transitions remains.\cite{tubule_unpublished}

\section{Summary}
\label{summary_section}

In these notes, I have presented a small cross-section of the beauty
and richness of fluctuating polymerized membranes. I have demonstrated
the importance of the in-plane order in determining the long-scale
conformations of these elastic sheets, by discussing in-plane
anisotropy and local random heterogeneity and showed that these lead
to a rich and highly nontrivial phenomenology.

\section{Acknowledgments}
 
\vskip .2in 

The material presented in these lectures is an outcome of research done
with a number of wonderful colleagues. The physics of anisotropic membranes
presented in Section 1 is primarily based on extensive work done 
with John Toner. The Section 2 on heterogeneities in polymerized
membranes is based on many years of fruitful
collaboration with David Nelson and Pierre Le Doussal. And the
final section is based on a collaboration with Xiangjun
Xing, Tom Lubensky and Ranjan Mukhopadhyay. I am indebted
to these colleagues for much of my insight into the material presented
here. This work was supported by the National
Science Foundation through grants DMR-0321848 and MRSEC DMR-0213918,
the Lucile and David Packard Foundation and the A. P. Sloan Foundation.


\begin{thebibliography}{000}
\bibitem{persistent_length} Crumpling, however, only takes place
  beyond an exponentially long persistence length $\xi_p =  
  a e^{4\pi\kappa/3k_B T}$, that, in a typical liquid membrane at room
  temperature far exceeds its size, and therefore even a liquid membrane
  appears not crumpled.
\bibitem{notGeneric} Actually, in most current experimental
  realizations, the polymerization is random and leads to an isotropic
  membrane, making the isotropic case quite generic.
\bibitem{EfieldComment} In these lectures we focus on the anisotropy
  in the membrane polymer network. This contrasts strongly with
  the anisotropy of the embedding space that membrane is fluctuating
  in.  In this latter very interesting case 
  considered in the work of Tokuyasu and Toner\cite{TT} (with anisotropy
  introduced by application of an electric field), anisotropy has
  important qualitative effects even in the flat phase, where it
  lowers the lower-critical dimension for elastic nonlinearities down
  to $D_{uc}=5/2$ and leads to a new anomalous elasticity fixed point with
  new universal exponents.
\bibitem{commentGaussianFP} For example, spin waves in a ferromagnetic
  state or antiferromagnetic state and phonons in an ordinary crystal
  (even in 2d), at long scales are faithfully described by a harmonic
  Hamiltonian, controlled by a Gaussian fixed point.
\bibitem{largeNcomment} This is analogous to the familiar $1/n$ expansion
for critical phenomena, in which one expands about the number of spin
components $n\rightarrow\infty$ limit.
\bibitem{commentSoft} We discuss here only ``ideal'' gels that exhibit a
   statistically isotropic and homogeneous high-temperature phase, in
   contrast to the so-called ``semisoft'' gels that are intrinsically
   anisotropic and therefore do not exhibit true nemato-elastic
   Goldstone mode.
\bibitem{phi4comment} The situation is qualitatively similar to that
 of the $\phi^4$ theory describing Ising transition. There too, the
 dominant nonlinearity is $\phi^4$ that becomes relevant for $D<4$,
 with the subdominant $\phi^6$ nonlinearity near the Gaussian fixed
 point turning on only for $D\leq 3$. Despite of this one expects and
 indeed finds that $\phi^6$ operator can be neglected even for $D=3$
 and $D=2$, as the relevance analysis needs to be done near the
 interacting (Wilson-Fisher) and not near the Gaussian critical
 point. In the same way, we expect that subdominance of in-plane
 nonlinearities will survive even below $D=3$, but probably not all
 the way down to the physical case of $D=2$.
\end{thebibliography}
\end{document}